\documentclass[submission,copyright,creativecommons]{eptcs}

\usepackage{amsmath}
\usepackage{amssymb}
\usepackage{bookmark}
\usepackage{underscore}
\usepackage{graphicx}
\usepackage{adjustbox}
\usepackage{array}
\usepackage{booktabs}
\usepackage{caption}    
\usepackage{subcaption} 
\usepackage{wrapfig}
\usepackage{multirow}
\usepackage{xcolor}

\usepackage{hyperref}
\usepackage[nameinlink,noabbrev]{cleveref}

\usepackage[inline]{enumitem}
\newcommand{\inliststyle}[1]{\textbf{\small #1}}
\newlist{inlist}{enumerate*}{1}
\setlist[inlist]{label={\inliststyle{(\arabic*)}}}

\title{SyGuS-Comp 2018: Results and Analysis}
\author{
	Rajeev Alur 
	\institute{University of Pennsylvania, USA}
	\email{alur@cis.upenn.edu}
  \and
	Dana Fisman
	\institute{Ben-Gurion University, Israel}
	\email{dana@cs.bgu.ac.il}
  \and
	Saswat Padhi
	\institute{University of California, \\ Los Angeles, USA}
	\email{padhi@cs.ucla.edu}
  \and
	Rishabh Singh
	\institute{Google Brain, USA}
	\email{rising@google.com}
  \and
	Abhishek Udupa
	\institute{Microsoft, USA}
	\email{abudup@microsoft.com}
}

\def\titlerunning{SyGuS-Comp 2018: Results and Analysis}
\def\authorrunning{R. Alur, D. Fisman, S. Padhi, R. Singh \& A. Udupa}
\def\copyrightholders{Alur, Fisman, Padhi, Singh \& Udupa}
\begin{document}
\maketitle


\newcommand{\alc}{\textsc{Alchemist-cs}}
\newcommand{\alccsdt}{\textsc{Alchemist-csdt}}
\newcommand{\cvclast}{\ensuremath{\textsc{CVC4}_{2017}}}
\newcommand{\enum}{\textsc{Enumerative}}
\newcommand{\skac}{\textsc{Sketch-ac}}
\newcommand{\ice}{\textsc{Ice-dt}}
\newcommand{\toast}{\textsc{SosyToast}}
\newcommand{\stoch}{\textsc{Stochastic}}
\newcommand{\eusolverlast}{\ensuremath{\textsc{EUSolver}_{2016}}}
\newcommand{\euphony}{\textsc{Euphony}}
\newcommand{\ethree}{\textsc{e3solver}}

\newcommand{\cvc}{\textsc{CVC4}}
\newcommand{\horndini}{\textsc{Horndini}}
\newcommand{\cvcnew}{\ensuremath{\textsc{CVC4}_{2018}}}
\newcommand{\eusolvernew}{\ensuremath{\textsc{EUSolver}_{2017}}}
\newcommand{\eusolver}{\textsc{EUSolver}}
\newcommand{\dryd}{\textsc{DryadSynth}}
\newcommand{\lig}{\textsc{LoopInvGen}}
\newcommand{\sygus}{SyGuS}
\newcommand{\comp}{SyGuS-Comp}

\newcolumntype{R}[2]{%
  >{\adjustbox{angle=#1,lap=\width-(#2)}\bgroup}%
  l%
  <{\egroup}%
}
\newcommand*\rot{\multicolumn{1}{R{90}{1em}}}

\newcommand{\verify}[1]{\textcolor{red}{#1}}


\begin{abstract}
    \emph{Syntax-guided synthesis (SyGuS)} is the computational problem of finding an implementation $f$
    that meets both a semantic constraint given by a logical formula $\varphi$ in a background theory $T$,
    and a syntactic constraint given by a grammar $G$,
    which specifies the allowed set of candidate implementations.
    Such a synthesis problem can be formally defined in the \emph{SyGuS input format (SyGuS-IF)},
    a language that is built on top of SMT-LIB.

    The \emph{syntax-guided synthesis competition (\comp)} is an
    effort to facilitate, bring together and accelerate research and development of efficient
    solvers for SyGuS by providing a platform for evaluating different synthesis
    techniques on a comprehensive set of benchmarks. 
    In the 5\textsuperscript{th} \comp{}, five solvers competed on over 1600 benchmarks across various tracks.
    This paper presents and analyses the results of this year's (2018) SyGuS competition.
\end{abstract}

\section{Introduction}
\label{sec:intro}

The \emph{Syntax-Guided Synthesis} competition (\comp) is an annual competition aimed to provide
an objective platform for comparing different approaches for solving the syntax-guided synthesis problem.
A SyGuS problem takes as input a logical specification $\varphi$ for what a synthesized function $f$ should compute,
and a grammar $G$ providing syntactic restrictions on the implementation for the function $f$ to be synthesized.
Formally, a solution to a SyGuS instance $(\varphi,G,f)$ is a function $f_{imp}$ that is expressible in the grammar $G$
such that the formula $\varphi[f/f_{imp}]$ obtained on replacing $f$ by $f_{imp}$ in the logical specification $\varphi$ is valid.
SyGuS instances are formulated in SyGuS-IF~\cite{RaghothamanU14}, a format built on top of SMT-LIB2~\cite{smtlib}.

We report here on the 5\textsuperscript{th} SyGuS competition that took place in July 2018,
in Oxford, UK as a satellite event of CAV'18 (The 30\textsuperscript{th} International Conference on Computer-Aided Verification)
and SYNT'18 (The 7\textsuperscript{th} Workshop on Synthesis).
As in the previous year's competition, there were four tracks:
the general track, the conditional linear integer arithmetic (CLIA) track, the invariant synthesis (Inv) track,
and the programming by examples (PBE) track.
We assume that most readers of this report would already be familiar with the SyGuS problem and the \comp{} tracks,
and thus refer unfamiliar readers to the report on last year's competition~\cite{SyGuSComp17}.

The report is organized as follows.
\begin{itemize}[topsep=0.25em]
    \item \Cref{sec:participation} briefly describes the benchmarks and the key idea behind the submitted solvers.
    \item \Cref{sec:exp-set} provides details on the experimental setup.
    \item \Cref{sec:comp-results} gives an overview of the results per track.
    \item \Cref{sec:benchs-pres} provides details on the results, given from a single benchmark perspective.
    \item \Cref{sec:discussion} concludes the report with some key takeaway points.
\end{itemize}
 
\section{Submitted Benchmarks and Solvers}
\label{sec:participation}

In addition to the benchmarks from the last year's competition,
we received over $100$ new benchmarks this year,
across various competition tracks, which we summarize below in \Cref{tbl:new-benchmarks}.

\begin{table}[!h]
	\def\arraystretch{1.1}
	\small
	\begin{center}
		\begin{tabular}{r||c|l}
			\textbf{Track} & \textbf{Benchmarks} 	& \textbf{Contributors} \\\hline \hline
			CLIA 		   & 	15 		   			& Kangjing Huang (Purdue University) \\
			General 	   & 	29 		   			& Qinheping Hu and Loris D' Antoni (University of Wisconsin-Madison) \\
			Invariants 	   & 	21 + 32    			& Saswat Padhi (UCLA) $\enskip$+$\enskip$ Kangjing Huang (Purdue University) \\
			PBE-Strings    & 	10 		   			& Woosuk Lee (University of Pennsylvania) \\
		\end{tabular}
	\end{center}
	\captionsetup{skip=0em}
	\caption{New benchmarks contributed to various tracks}
	\label{tbl:new-benchmarks}
\end{table}

Five solvers were submitted to this year's competition:
\begin{inlist}
	\item \cvcnew, an improved version of \cvc,
	\item \dryd, a solver specialized for conditional linear integer arithmetic,
	\item \eusolvernew, an improved version of \eusolver,
	\item \horndini, a solver specialized for constrained horn clauses (CHCs) and
	\item \lig, a solver specialized for invariant generation problems.
\end{inlist}
\Cref{tbl:solvers-authors} lists the submitted solvers along with their authors,
and \Cref{tbl:solvers-in-tracks} shows the tracks in which each solver participated.

\begin{table}[b]
	\small
	\begin{center}
		\scalebox{0.975}{
		\renewcommand{\arraystretch}{1.2}
		\begin{tabular}{r||l}
			\textbf{Solver} & \textbf{Authors} \\ \hline \hline
			\cvcnew 		& Andrew Reynolds (Univ. of Iowa), Haniel Barbosa (Univ. of Iowa), Andrez Notzli (Stanford), \\
							& Cesare Tinelli (Univ. of Iowa), and Clark Barrett (Stanford) \\[6pt]
			\dryd           & KangJing Huang (Purdue Univ.), Xiaokang Qiu (Purdue Univ.), Qi Tan (Nanjing Univ.), and \\
							& Yanjun Wang (Purdue Univ.) \\[6pt]
			\eusolvernew  	& Arjun Radhakrishna (Microsoft) and Abhishek Udupa (Microsoft) \\[6pt]
			\horndini  		& Deepak D’Souza (IISc, Bangalore), P. Ezudheen (IISc, Bangalore), P. Madhusudan (UIUC), \\
							& Pranav Garg (Amazon), Daniel Neider (MPI-SWS), and Shubham Ugare (IIT, Guwahati) \\[6pt]
			\lig            & Saswat Padhi (UCLA), Rahul Sharma (Microsoft Research), and Todd Millstein (UCLA) \\
		\end{tabular}}
	\end{center}
	\captionsetup{skip=0em}
	\caption{List of registered solvers}
	\label{tbl:solvers-authors}
\end{table}

The \cvcnew\ solver is based on an approach for program synthesis that is implemented inside an SMT solver~\cite{ReynoldsDKTB15}.
This approach extracts solution functions from unsatisfiability proofs of the negated form of synthesis conjectures,
and uses counterexample-guided techniques for quantifier instantiation (CEGQI) that make finding such proofs practically feasible.
\cvcnew\ also combines enumerative techniques, and symmetry breaking techniques~\cite{ReynoldsT17}. 

The \dryd\ solver combines enumerative and symbolic techniques.
It considers benchmarks in conditional linear integer arithmetic theory (LIA), and can therefore assume all have a solution in some pre-defined decision tree normal form.
It then tries to first get the correct height of a normal form decision tree, and then tries to synthesize a solution of that height.
It makes use of parallelization, using as many cores as are available, and of optimizations based on solutions of typical LIA SyGuS problems.

The \eusolvernew\ solver uses a divide-and-conquer strategy~\cite{AlurCAV15} to find different expressions that satisfy different subsets of the input space,
and then unifies them into a solution that works well for the entire space of inputs.
Subexpressions are typically found using enumeration techniques
and are then unified into the final solutions using decision tree learning~\cite{AlurRU17}.

\begin{wraptable}{r}{5cm}
	\setlength{\tabcolsep}{4pt}
	\def\arraystretch{1.2}
	\begin{center}
		\begin{tabular}{r||rrrrr}
			& \multicolumn{5}{c}{\textbf{Solver}} \\[8pt]
			\textbf{Track} & \rot{\cvcnew} & \rot{\dryd} & \rot{\eusolvernew} & \rot{\horndini} & \rot{\lig} \\ \hline \hline
			CLIA        & 1 & 1 & 1 & 0 & 0 \\
			INV         & 1 & 1 & 1 & 1 & 1 \\
			General     & 1 & 0 & 1 & 0 & 0 \\ 
			PBE-Strings & 1 & 0 & 1 & 0 & 0 \\ 
			PBE-BV      & 1 & 0 & 1 & 0 & 0 \\
		\end{tabular}
	\end{center}
	\captionsetup{skip=0em}
	\caption{Participating solvers}
	\label{tbl:solvers-in-tracks}
\end{wraptable}

The \horndini\ solver extends the classical IC3 decision-tree algorithm with
the Horn implication counterexamples (Horn-ICE) framework~\cite{EzudheenND0M18},
which extends the ICE-learning model.
The authors describe a decision-tree learning algorithm that learns from Horn-ICE samples,
works in polynomial time, and uses statistical heuristics to learn small trees that satisfy the samples.

The \lig\ solver~\cite{PadhiM17} for invariant synthesis extends the data-driven approach to inferring sufficient loop invariants
from a set of program states~\cite{PadhiSM16}.
Previous approaches to invariant synthesis were restricted to using a fixed set, or a fixed template for features,
e.g., ICE-DT~\cite{GNMR16} requires the shape of constraints (such as octagonal) to be fixed apriori.
Instead \lig\ starts with no initial features, and automatically learns features as necessary using program synthesis.
It reduces the problem of loop invariant inference to a series of precondition inference problems,
and uses a counterexample-guided inductive synthesis (CEGIS) loop to revise the current candidate.
\section{Experimental Setup} 
\label{sec:exp-set}

The solvers were run on the StarExec platform~\cite{starexec} with a dedicated cluster of 12 nodes,
where each node consisted of two 4-core 2.4\,GHz Intel processors with 256\,GB RAM and 1\,TB hard-disk space.
The memory usage limit for each solver run was set to 128\,GB, and the wall-clock time limit is set to 3600 seconds
(thus, a solver that used all 4 cores could consume at most 14400 seconds of CPU time).
The solutions that the solvers produced were checked for both syntactic and semantic correctness.
That is, a first postprocessor checked that the produced expression adhered to the grammar specified in the given benchmark,
and if this check passes, a second postprocessor checked that the solution adhered to semantic constraints
given in the benchmark (by invoking an SMT solver).
\section{Results Overview}
\label{sec:comp-results}



The primary criterion for winning a track was the number of benchmarks solved,
but we also analyzed the time to solve and the the size of the generated expressions.
The overall score for each solver was computed as $5N + 3F + S$.
Here $N$ denotes the number of benchmarks solved by the solver,
$F$ denotes the number of benchmarks solved among the fastest,
and $S$ denotes the number of benchmarks for which the size of the generated solution was among the shortest.
We used a pseudo-logarithmic scale for $F$ and $S$.
For time to solve, the scale is: $[0,1)$, $[1,3)$, $[3,10)$, $[10,30)$, $[30, 100)$,
$[100,300)$, $[300, 1000)$, $[1000,3600)$, $\geqslant 3600$.
That is, the first ``bucket'' refers to termination in less than one second,
the second to termination in one to three seconds and so on.
We say that a solver solved a certain benchmark \emph{among the fastest}
if the time it took to solve that benchmark is in the same bucket
as that of the solver which solved that benchmark in minimum time.
Similarly, for expression sizes, the pseudo-logarithmic scale we use is:
$[1,10)$, $[10,30)$, $[30,100)$, $[100,300)$, $[300,1000)$, $\geqslant 1000$,
where expression size is the number of nodes in the SyGuS parse-tree.
We also report on the number of benchmarks \emph{solved uniquely} by a solver,
\emph{i.e.} the number of benchmarks which no solver other than the particular solver could solve.

\begin{figure}
	\begin{center}
		\vspace{3em}
		\begin{minipage}{\textwidth}
			\centering%
			\includegraphics[width=0.5\textwidth]{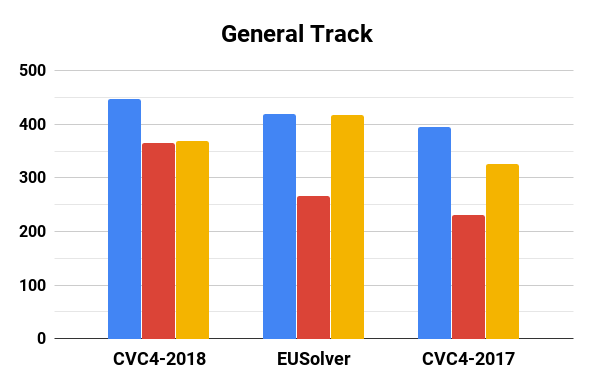}%
			\includegraphics[width=0.5\textwidth]{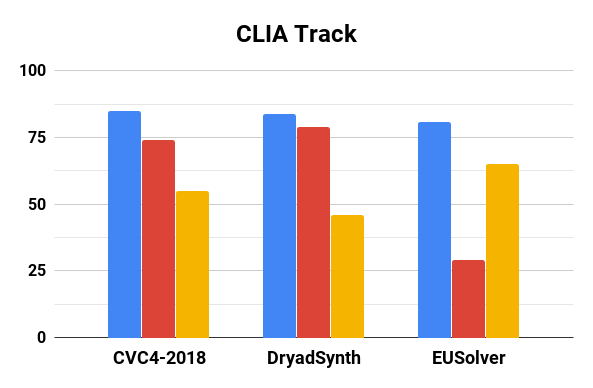}
		\end{minipage}
		\\[1cm]
		\begin{minipage}{\textwidth}
			\centering%
			\includegraphics[width=0.5\textwidth]{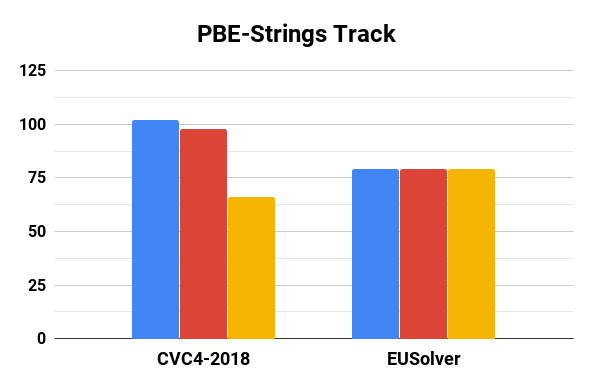}%
			\includegraphics[width=0.5\textwidth]{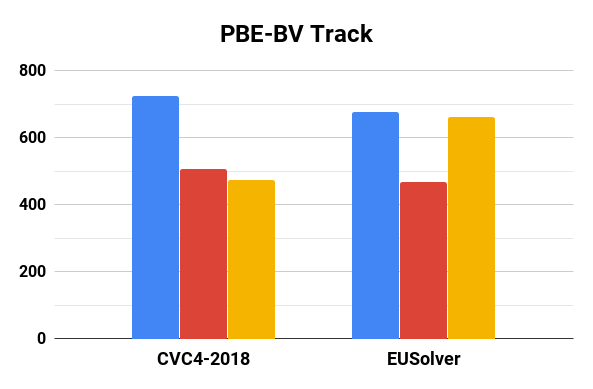}
		\end{minipage}
		\\[1cm]
		\begin{minipage}{\textwidth}
			\centering%
			\includegraphics[width=\textwidth]{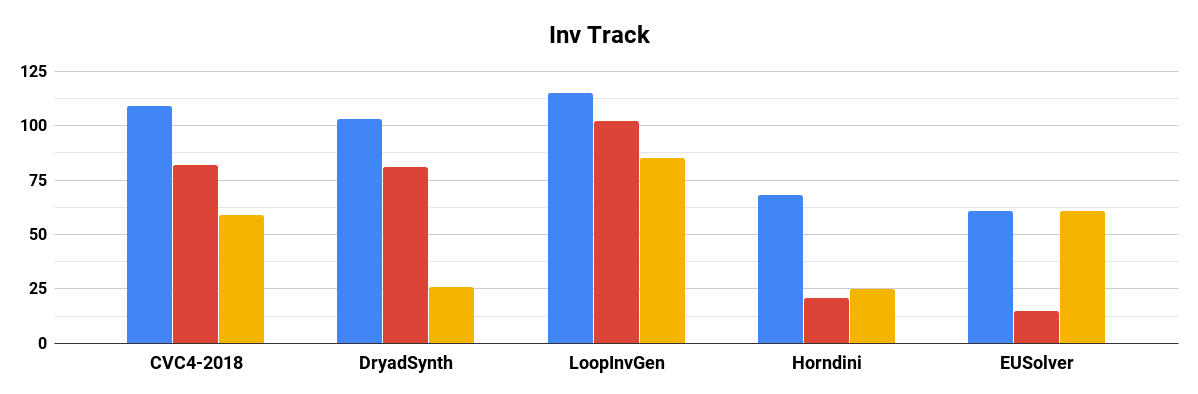}\\[0.5em]
			\includegraphics[width=0.75\textwidth]{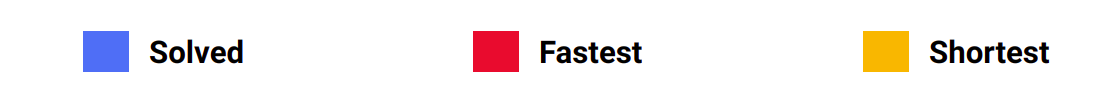}
		\end{minipage}
	\end{center}
	\caption{The number of benchmarks solved by different solvers across all tracks,
			 the number of benchmarks a solver solved among the fastest,
			 and the number of benchmarks for which a solver generated an expression among the smallest size.}
	\label{fig:resultsPerTrack}	
\end{figure}

In \Cref{fig:resultsPerTrack}, we show the number of benchmarks solved,
the number of benchmarks solved among the fastest,
and the number of synthesized expressions among the smallest size;
per solver per track.

\paragraph{General Track}
In the general track, \cvcnew\ solved the most number of benchmarks ($448$),
and \eusolvernew\ came second, solving $420$.
We note that the new version \cvcnew\ is significantly better than the previous version \cvclast,
which could only solve $398$ benchmarks.
The same order appears in the number of benchmarks solved among the fastest:
\cvcnew\ with $366$, \eusolvernew\ with $266$, and \cvclast\ with $252$.
Finally, we note that \cvcnew\ is able to solve $12$ benchmarks that no other solver could solve,
and similarly there are $9$ benchmarks that only \eusolver\ could solve.

\begin{table}[t]
	\begin{center}
		\scalebox{0.85}{
		\def\arraystretch{1.125}
		\begin{tabular}{lr||rrrrrrrrrrrr|r}
                                          &	 & \rot{Compiler Optimizations and Bit Vectors}
                                             & \rot{Let and Motion Planning}
                                             & \rot{Invariant Generation with Bounded Ints}
                                             & \rot{Invariant Generation with Unbounded Ints}
                                             & \rot{Multiple Functions}
                                             & \rot{Arrays}
                                             & \rot{Hackers Delight}
                                             & \rot{Integers}
                                             & \rot{Program Repair}
                                             & \rot{ICFP}
                                             & \rot{Cryptographic Circuits}
                                             & \rot{Instruction Selection}
                                             & \textbf{Total} \\\hline \hline
  \multicolumn{2}{r||}{Number of benchmarks} &  32  &   30  &   28  &   28  &   32  &   35  &   69  &   34  &   18  &   50  &   214 &   28  & 598 \\ \hline			 
  \multirow{3}{*}{\textbf{Solved}} & \cvcnew &	16	&   17	&   24	&   24  &	13	&   31	&   62  &	34  &	17  &	50  & 	160 & 	0   & 448 \\
		 			          & \eusolvernew &	16	&   10	&   24	&   23	&   18	&   31  &	53  & 	33  &	14  &	50  &	148 &	0   & 420 \\
 					              & \cvclast &  15	&   15  &   24  &	24  &	12	&   31	&   62	&   34	&   17	&   48	&   116 &	0   & 398 \\ \hline
 \multirow{3}{*}{\textbf{Fastest}} & \cvcnew &	15	&   15	&   22	&   24	&   9	&   31	&   59	&   33  &	16  &	23	&   119 &	0   & 366 \\
 								 & \eusolver &	13  &	1   &	12	&   11	&   14	&   5   &	29	&   15	&   12  & 	45	&   109	&   0   & 266 \\
								  & \cvclast &  12  &	9	&   16	&   14  &	9   &	24	&   60  &	33  &	6	&   20	&   49  & 	0   & 252 \\ \hline
\multirow{3}{*}{\textbf{Uniquely}} & \cvcnew &	1	&   2   & 	0	&   0   & 	0	&   0   & 	0	&   0   &	2	&   0	&   7   &	0   & 12 \\
								 & \eusolver &	3	&   0   &	0   &	0   &	6   &	0	&   0   &	0   &	0   &	0	&   0   &	0   & 9 \\
								  & \cvclast &  0	&   0	&   0   & 	0	&   0   & 	0	&   0   & 	0	&   0   & 	0   &	0	&   0   & 0 \\ \hline			
		\end{tabular}}
	\end{center}
	\captionsetup{skip=0em}
	\caption{The performance of various solvers across all categories of the general track}
	\label{tbl:general-categories}
\end{table}

We partitioned the benchmarks of the general track to a number of categories,
each containing a set of related benchmarks.
The results per category are given in the Table~\ref{tbl:general-categories}.
We observe that \eusolvernew\ preformed significantly better in the ``Multiple Functions'' and ``ICFP'' categories.
While the \cvcnew\ solver preformed better in the other categories,
none of the solvers could solve any of the benchmarks from the ``Instruction Selection'' category.


\paragraph{Conditional Linear Arithmetic Track}
In the CLIA track, \cvcnew\ and \dryd\ had a close competition.
\cvcnew\ solved $85$ out of $88$ benchmarks, \dryd\ solved $84$ benchmarks, and \eusolvernew\ solved $81$ benchmarks.
In terms of the time to solve, \dryd\ solved $79$ benchmarks among the fastest, \cvcnew\ solved $74$,
followed by \eusolvernew\ which solved $29$ among the fastest.
There were two benchmarks that were solved uniquely by \dryd,
and one that was solved uniquely by \cvcnew.

\paragraph{Invariant Generation Track}
In the invariant generation track, the \lig\ solver solved $115$ out of $127$ benchmarks, \cvcnew\ solved $109$,
\dryd\ solved $103$, \horndini\ solved $68$ and \eusolvernew\ solved $61$ benchmarks.
In terms of the time to solve, \lig\ solved $102$ benchmarks among the fastest, followed by \cvcnew\ which solved $82$,
\dryd\ which solved $81$, \horndini\ which solved $21$, and \eusolvernew\ which solved $15$.
There was one benchmark that was solved by a unique solver -- the \texttt{fib_17n.sl} benchmark solved by \lig. 
\vspace{-0.5em}
\paragraph{Programming By Example (Bit Vectors) Track}
In the PBE track on the theory of bit vectors, the \cvcnew\ solver solved $724$ out of $750$ benchmarks
and \eusolvernew\ solved $677$ benchmarks.
In terms of the time to solve, \cvcnew\ solved $508$ benchmarks among the fastest,
and \eusolvernew\ solved $468$.
However, \eusolvernew\ generates shorter expressions than \cvcnew\ in significantly many cases.
There were four benchmarks that were solved uniquely by \cvcnew,
and one benchmark that was solved uniquely by \eusolvernew.
\vspace{-0.5em}
\paragraph{Programming By Example (Strings) Track}
In the PBE track on the theory of strings, the \cvcnew\ solver solved $102$ out of $118$ benchmarks,
and \eusolvernew\ solved $79$ benchmarks.
In terms of the time to solve, \cvcnew\ solved $98$ benchmarks among the fastest,
and \eusolvernew\ solved $79$.
We note again that \eusolvernew\ generates shorter expressions than \cvcnew\ in several cases.
There were $21$ benchmarks that were solved uniquely by \cvcnew.
\section{Detailed Results}
\label{sec:benchs-pres}

In this section we show the results of the competition from the benchmark's perspective.
For a given benchmark we would like to know:
\begin{inlist}
	\item how many solvers solved it
	\item what are the minimum and maximum times required to solve
	\item what are the minimum and maximum sizes of solutions generated
	\item which solver solved the benchmark the fastest, and
	\item which solver produced the smallest expression.
\end{inlist}

\begin{figure*}
	\noindent\makebox[\textwidth]{
		\scalebox{0.625}{
			\begin{tabular}{c}
				\includegraphics[width=10in]{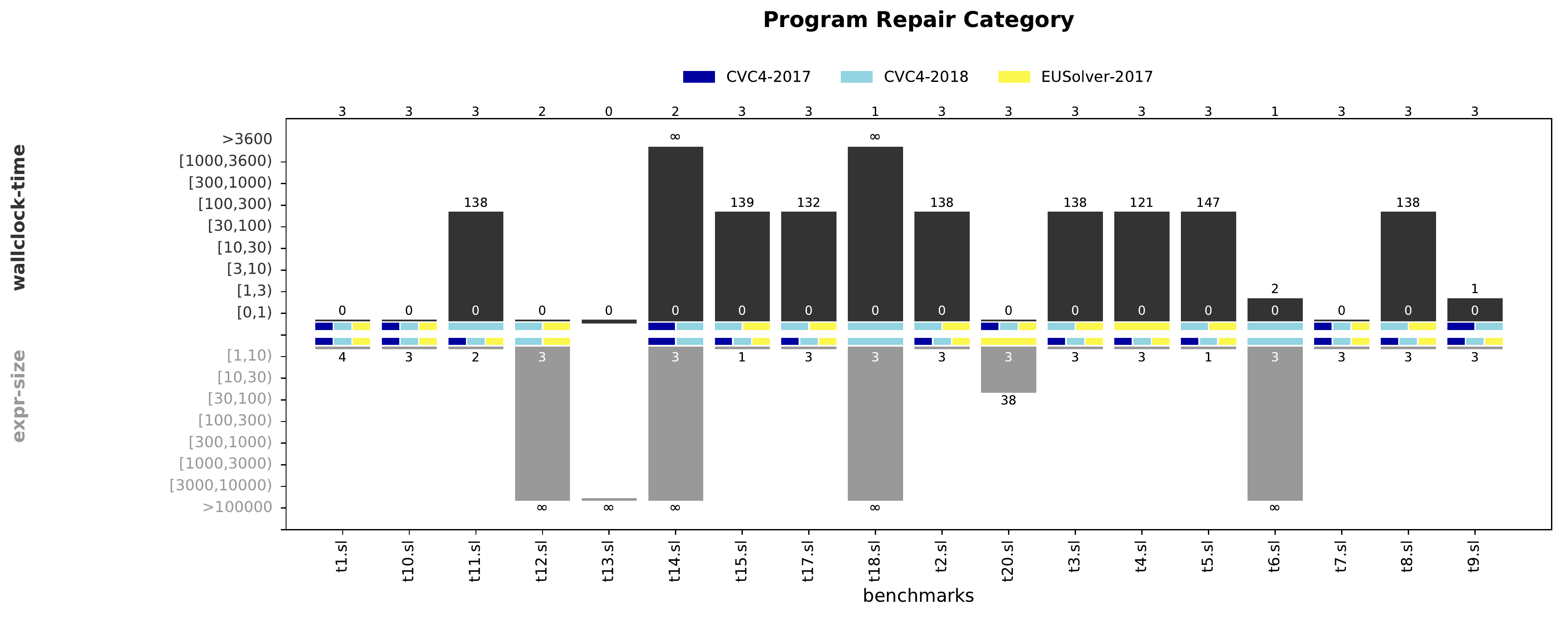} \\[3cm]
				\includegraphics[width=10in]{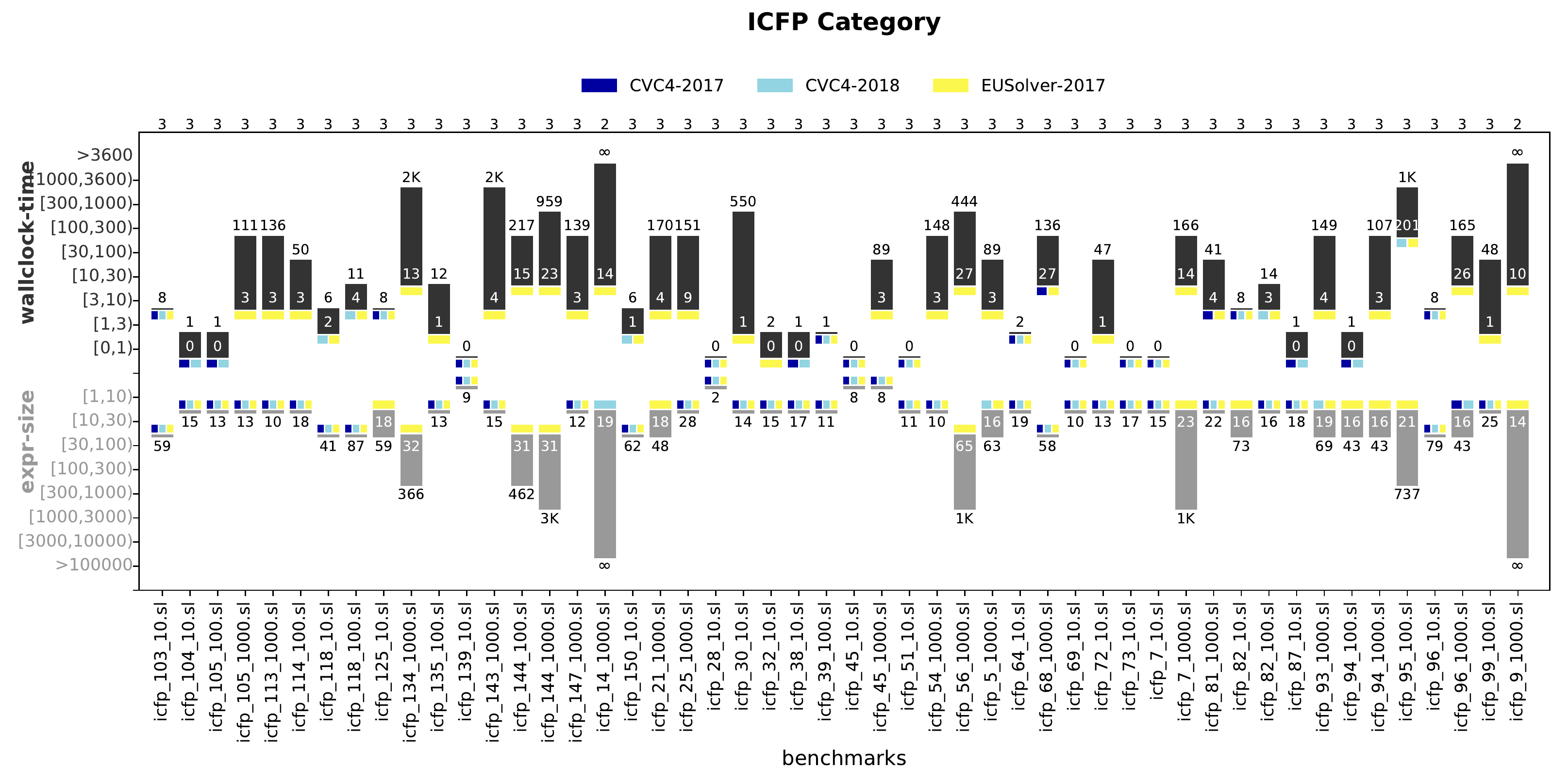}
			\end{tabular}
	}}
	\caption{Evaluation of program repair and ICFP categories of the General track.}
	\label{fig:prog-rep-icfp}
\end{figure*}

We present the results in groups organized per track and category.
For instance, the top plot in \Cref{fig:prog-rep-icfp} presents the details for program repair benchmarks from the general track.
The black bars above the $y$-axis show the range of time taken to solve across the various solvers, in our pseudo logarithmic scale.
Inspect for instance benchmark \texttt{t2.sl}.
The black bar indicates that the fastest solver takes less than $1$ second, and the slowest one takes between $100$ to $300$ seconds.
The black number above the black bar indicates the exact number of seconds (floor-rounded to the nearest second)
it took the slowest solver to solve a benchmark (and $\infty$ if at least one solver exceeded the time bound).
Thus, we can see that for \texttt{t2.sl}, the slowest solver took $138$ seconds.
The white number at the lower part of the bar indicates the time taken by the fastest solver.
Thus, we can see that for \texttt{t2.sl}, the fastest solver required less than $1$ second.
The colored squares/rectangles below the black bar indicate which solvers were among the fastest to solve that benchmark
(according to the solvers' legend at the top).
For instance, we can see that \cvcnew\ and \eusolvernew\ were the fastest to solve \texttt{t2.sl},
solving it in less than a second, and that among the solvers that solved \texttt{t4.sl} only \eusolvernew\ solved it in less than a second.

Similarly, the gray bars below the $y$-axis indicate the range of expression sizes in pseudo-logarithmic scales,
where the size of an expression is determined by the number of nodes in its parse tree.
The black number at the bottom of the gray bar indicates the exact size of the largest solution (or $\infty$ if it exceeded $1000$),
and the white number at the top of the gray bar indicates the exact size of the smallest solution.
When the smallest and largest size of expressions are in the same pseudo-logarithmic bucket, as is the case in \texttt{t2.sl}),
we indicate the expression size only in black.
The colored squares/rectangles above the gray bar indicate which solvers were amongst the ones that produced the smallest expression
(according to the solvers' legend at the top).
For instance, for \texttt{t20.sl} the smallest expression produced had size $3$,
which is produced only by \eusolvernew.

Finally, the top $x$-axis indicates the number of solvers that solved a particular benchmark.
For instance, in \Cref{fig:prog-rep-icfp}, only one solver solved \texttt{t6.sl}, two solvers solved \texttt{t14.sl},
three solvers solved \texttt{t2.sl}, and no solver solved \texttt{t13.sl}.
Note that for the benchmarks that no solver is able to solve, the black bars indicate the range of time taken by solvers to terminate.
When no solver produces a correct result, there are no colored squares/rectangles below the black bars, as is the case for \texttt{t13.sl}.

\begin{figure*}
	\noindent\makebox[\textwidth]{
		\scalebox{0.625}{
			\begin{tabular}{c}
				\includegraphics[width=10in]{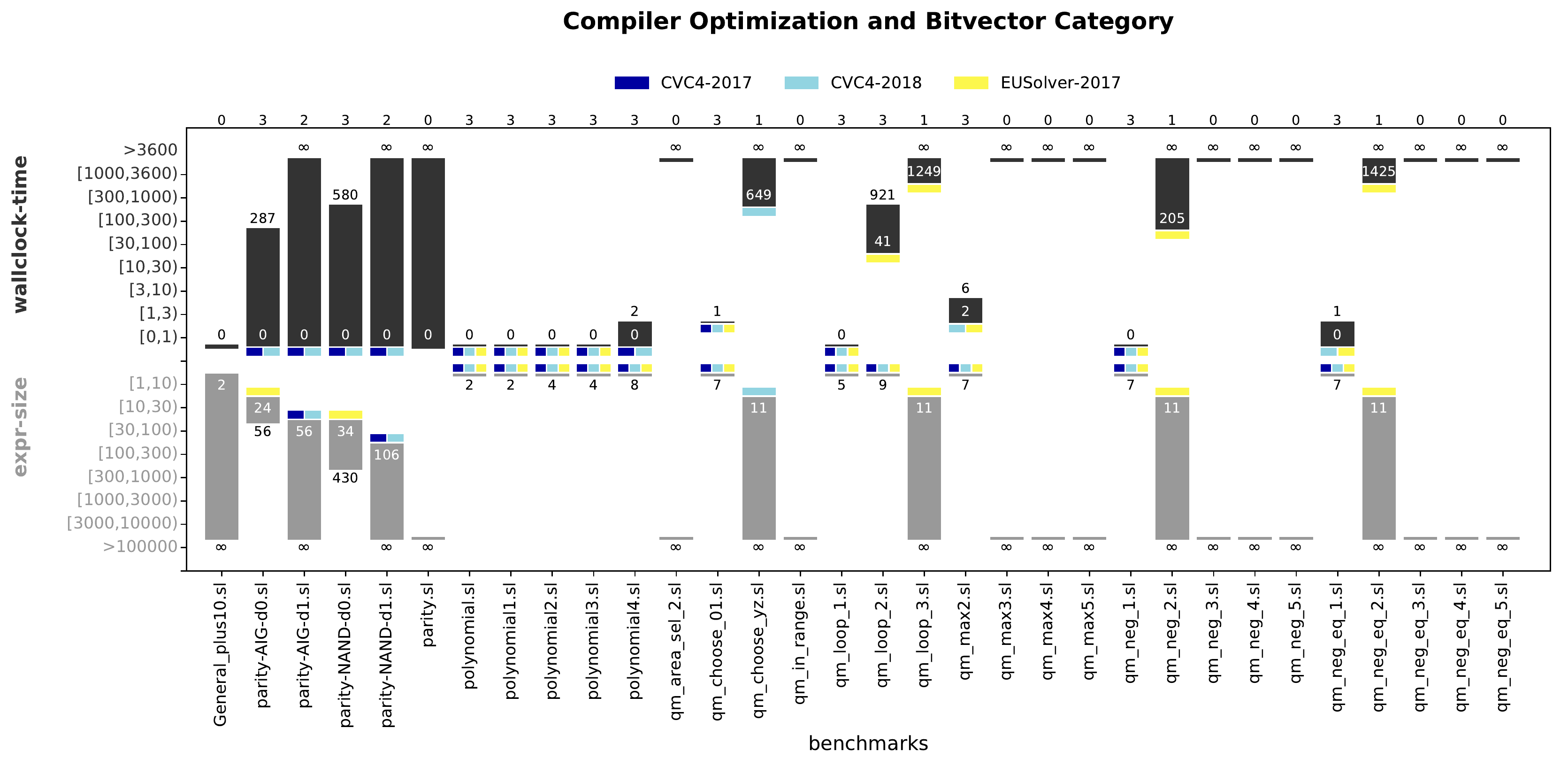} \\[3cm]
				\includegraphics[width=10in]{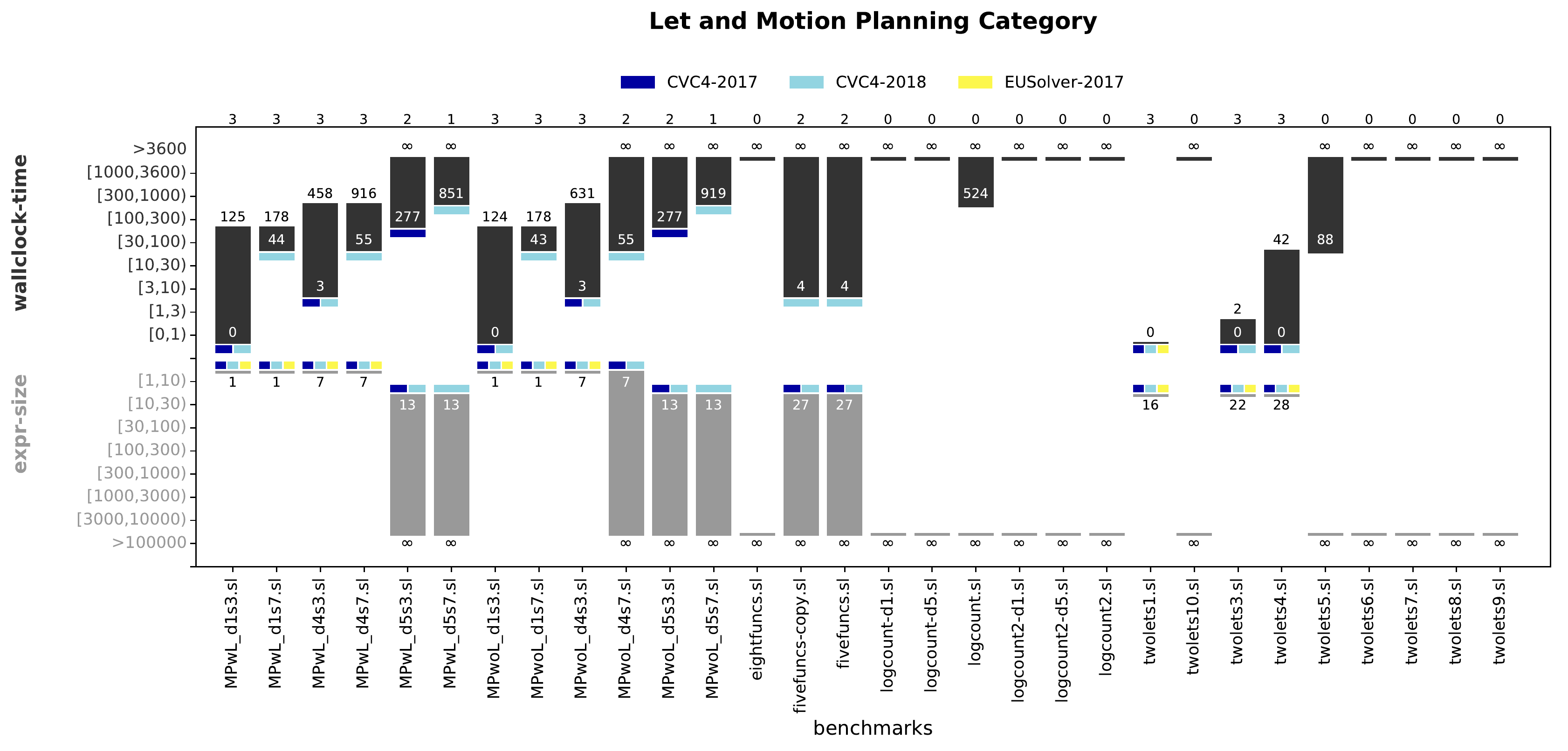}
			\end{tabular}
	}}
	\caption{Evaluation of compiler optimizations, bitvectors, let and motion planning,
			 and program repair categories of the General track.}
	\label{fig:let-mot-plan}
\end{figure*}

\begin{figure*}
	\noindent\makebox[\textwidth]{
		\scalebox{0.625}{
			\begin{tabular}{c}
				\includegraphics[width=10in]{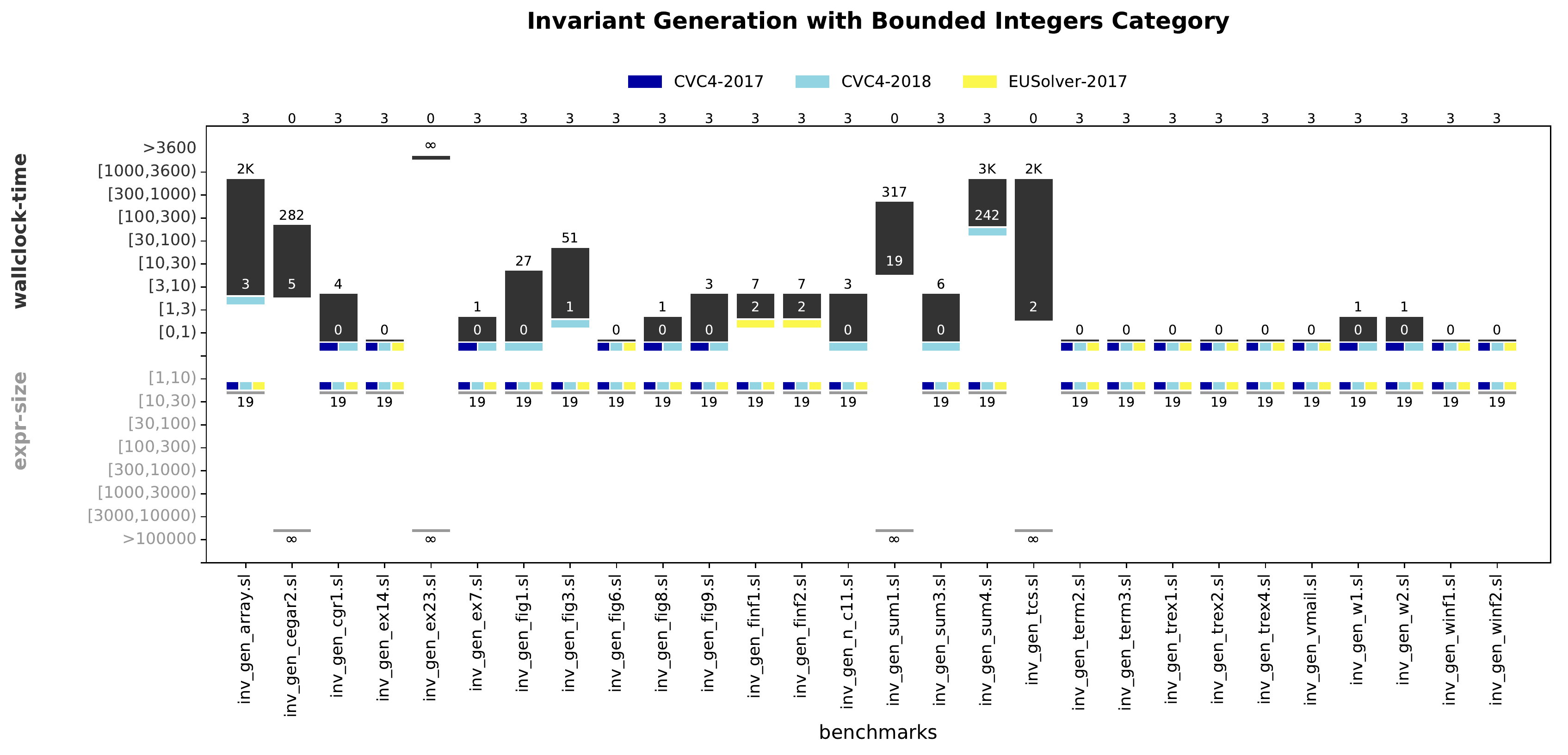} \\[3cm]
				\includegraphics[width=10in]{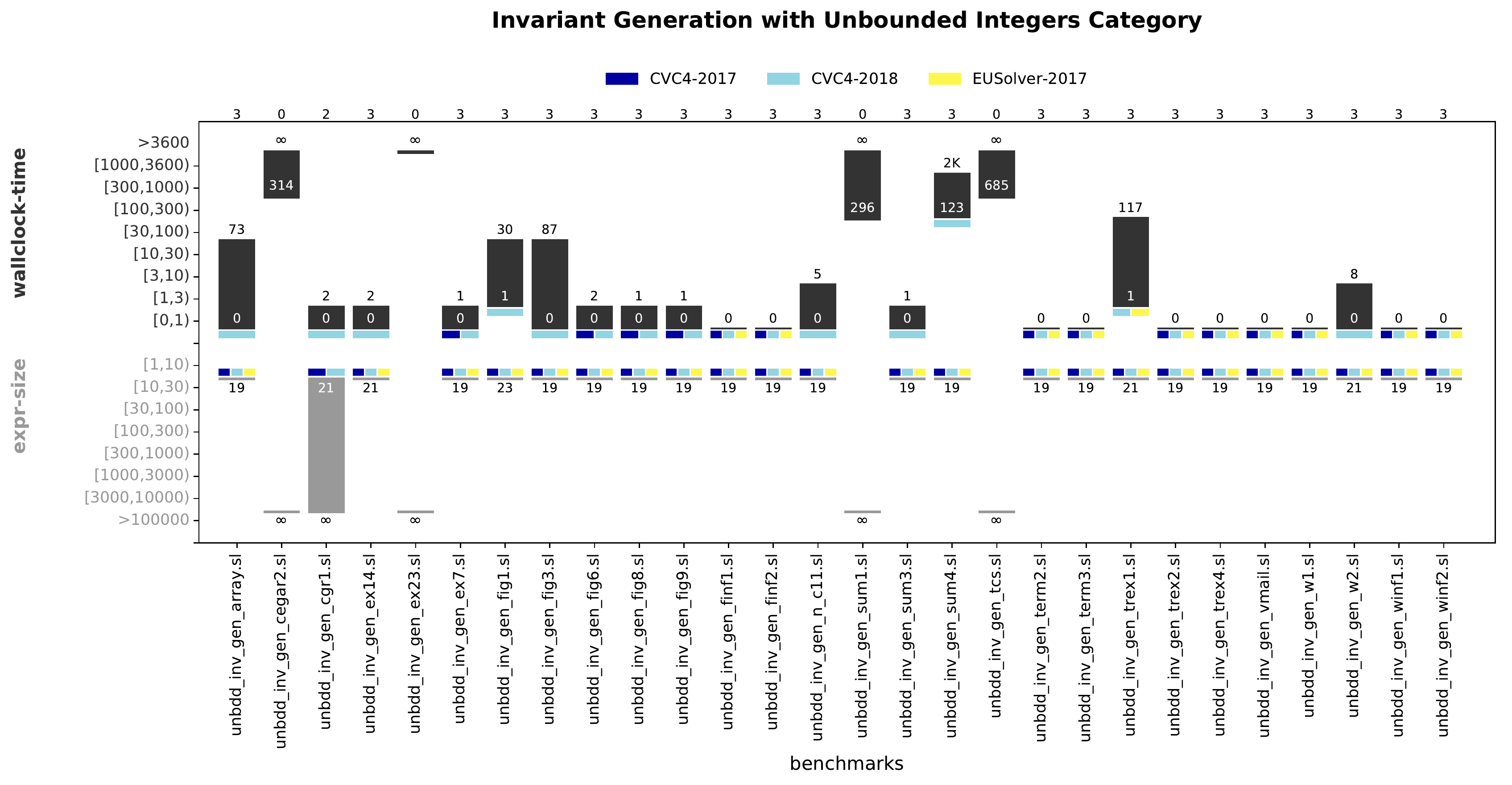} 
			\end{tabular}
	}}
	\caption{Evaluation of invariant generation categories of the General track.}
	\label{fig:inv-results}
\end{figure*}

\begin{figure*}
	\noindent\makebox[\textwidth]{
		\scalebox{0.625}{
			\begin{tabular}{c}
				\includegraphics[width=10in]{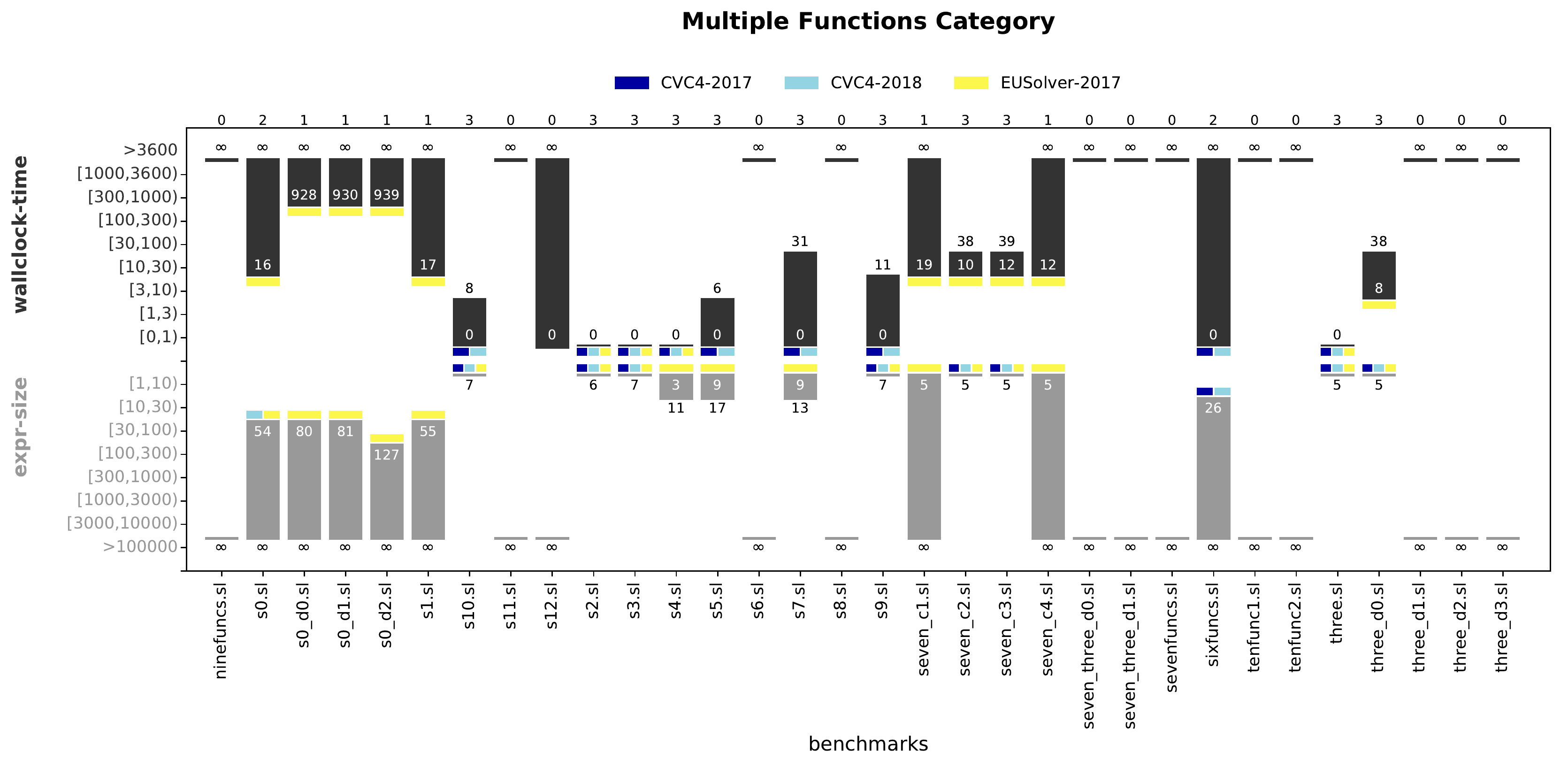} \\[3cm]
				\includegraphics[width=10in]{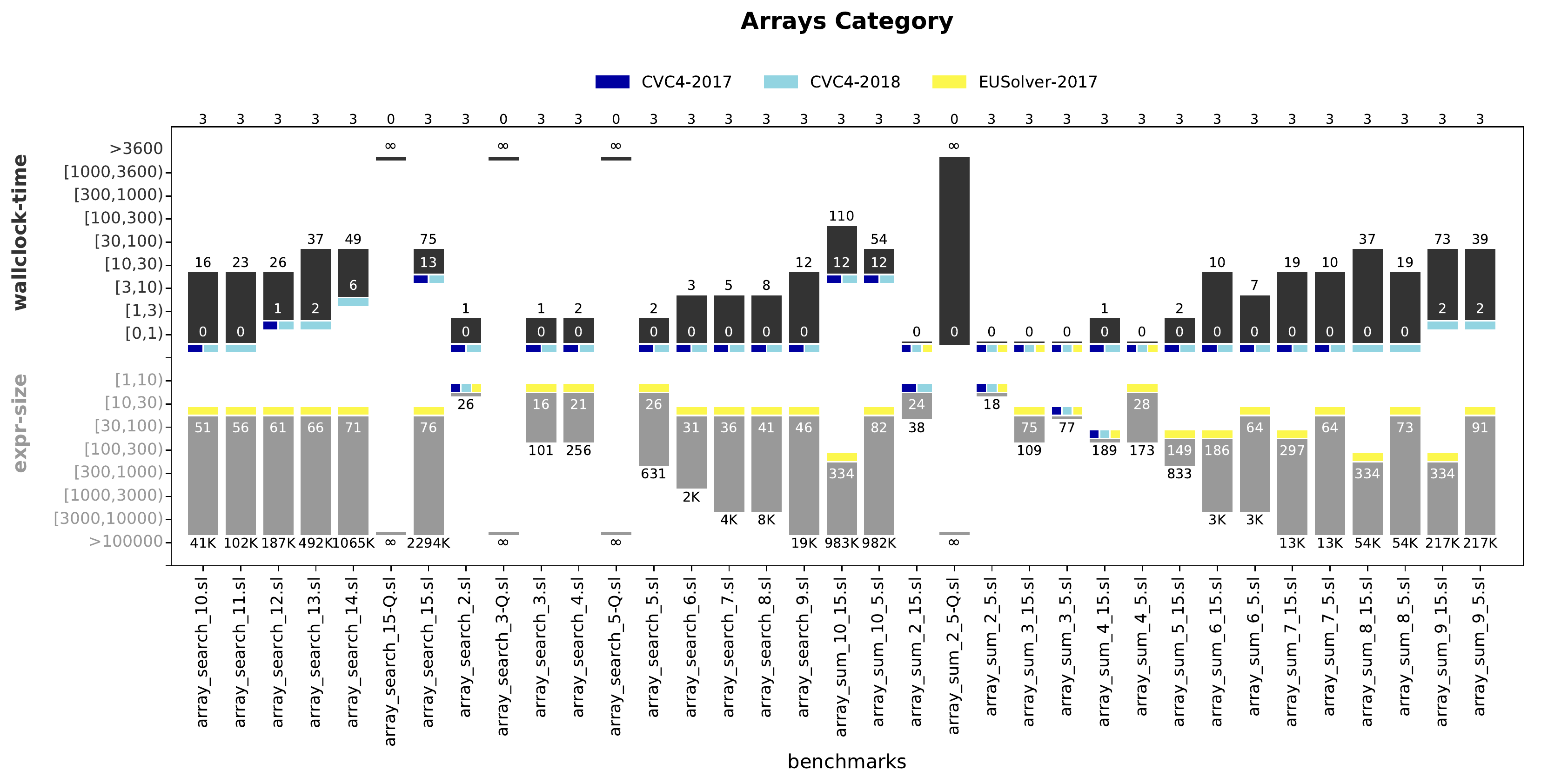} 
			\end{tabular}
	}}
	\caption{Evaluation of multiple functions and arrays categories of the General track.}
	\label{fig:mult-func-arr}
\end{figure*}

\begin{figure*}
	\noindent\makebox[\textwidth]{
		\scalebox{0.6}{
			\begin{tabular}{c}
				\includegraphics[width=10in]{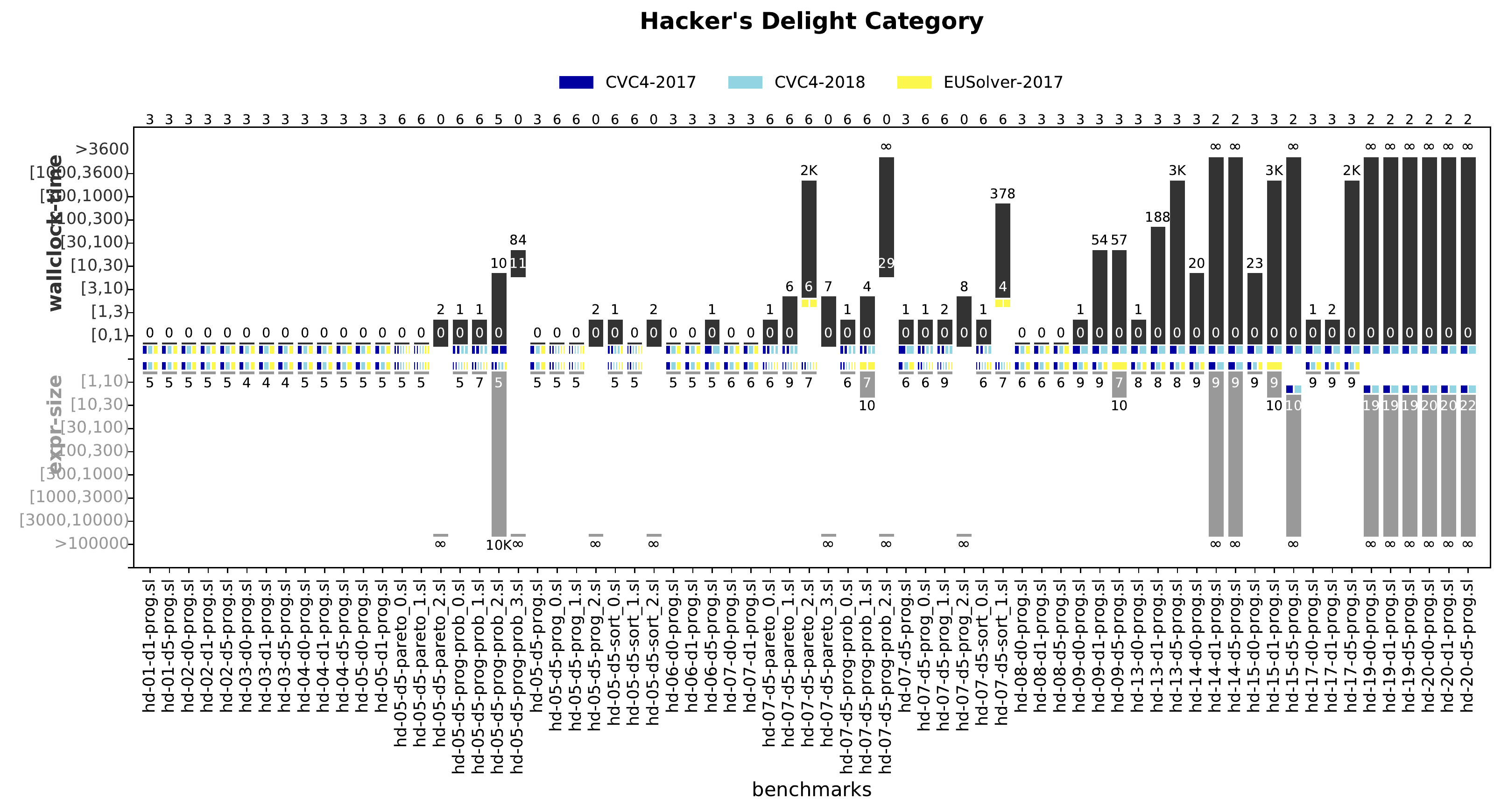} \\[3cm]
				\includegraphics[width=10in]{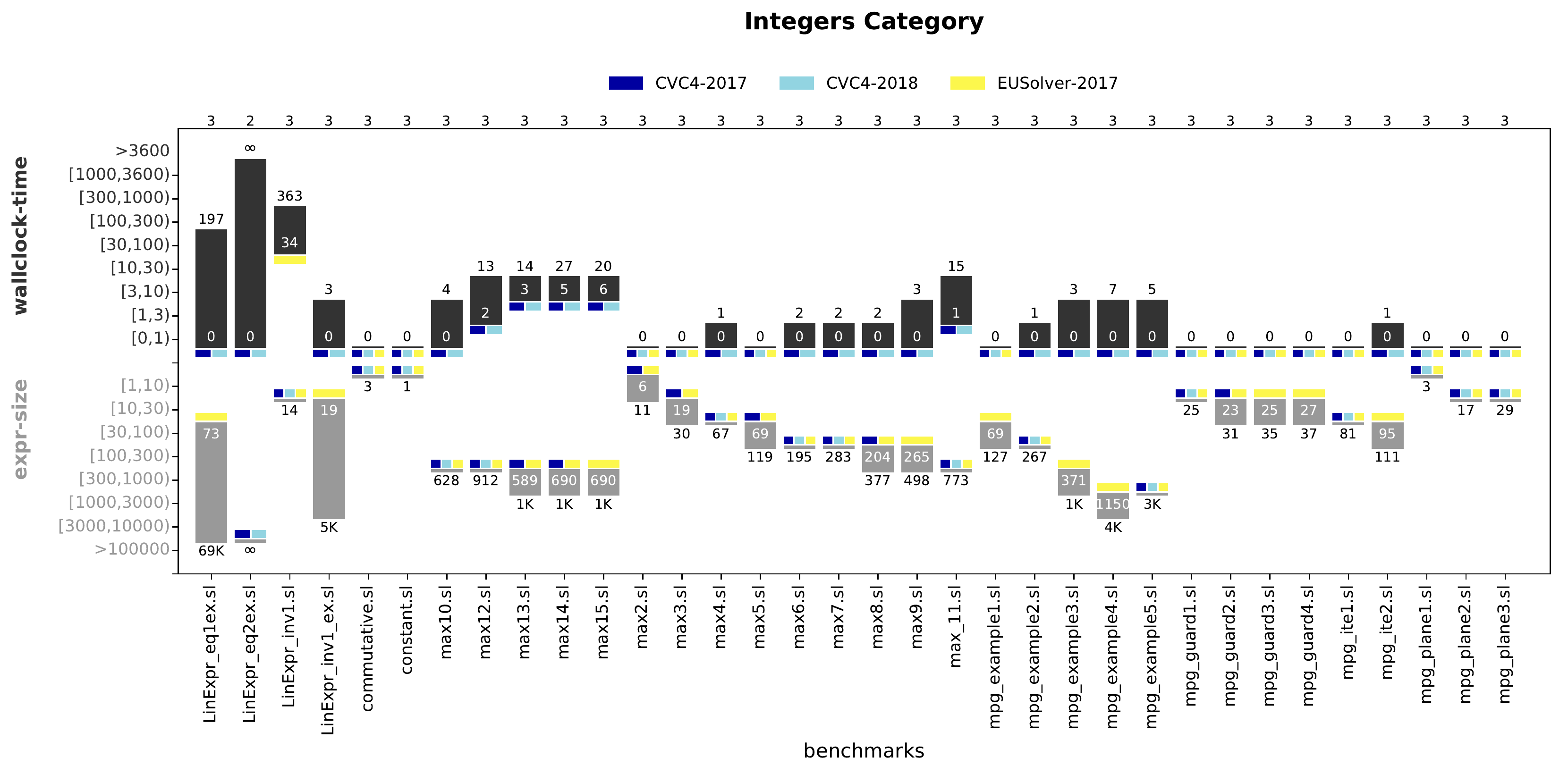} 
			\end{tabular}
	}}
	\caption{Evaluation of hacker's delight and integers categories of the General track.}
	\label{fig:hd-int}
\end{figure*}

\begin{figure*}
	\noindent\makebox[\textwidth]{
		\scalebox{0.625}{
			\begin{tabular}{c}
				\includegraphics[width=10in]{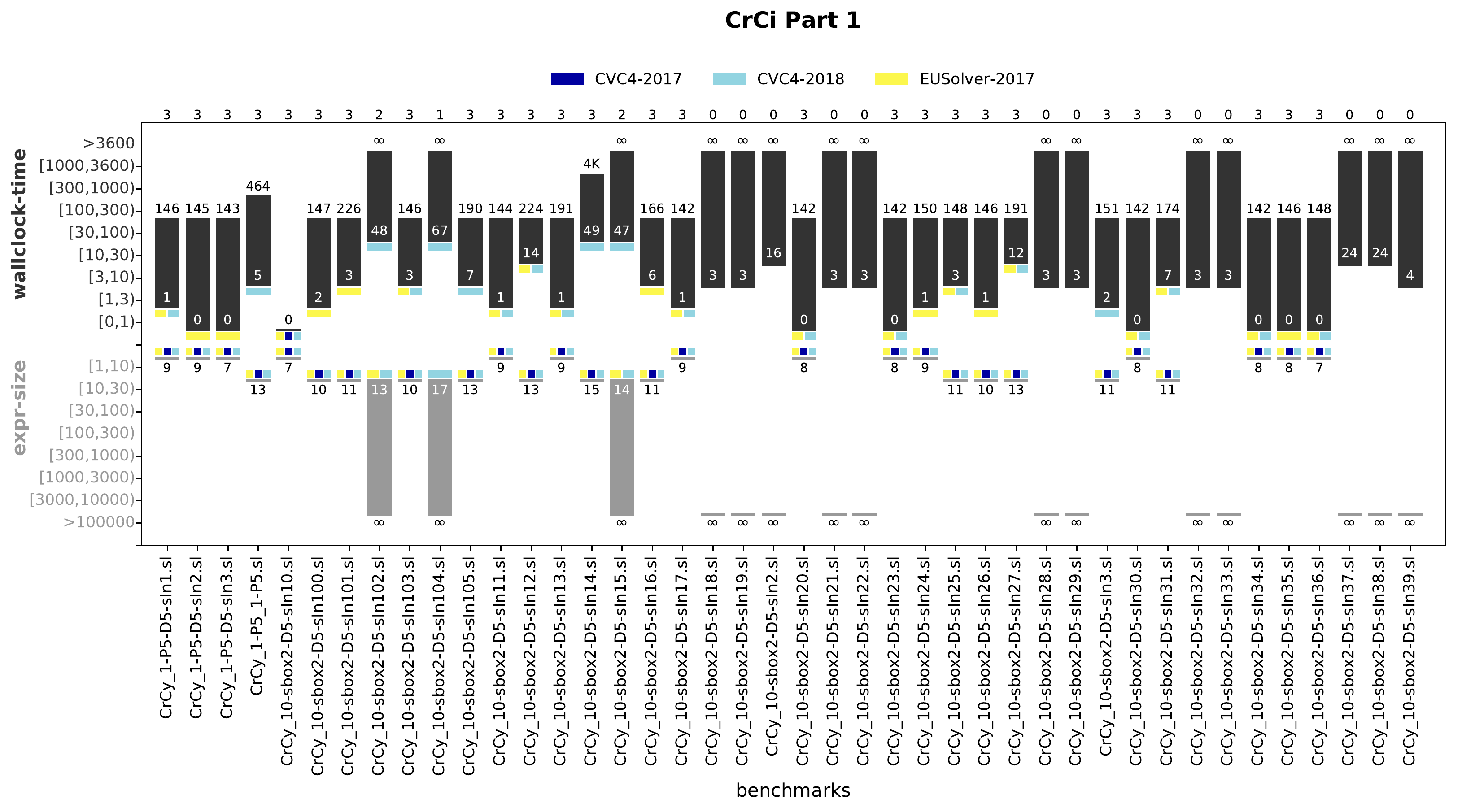} \\[3cm]
				\includegraphics[width=10in]{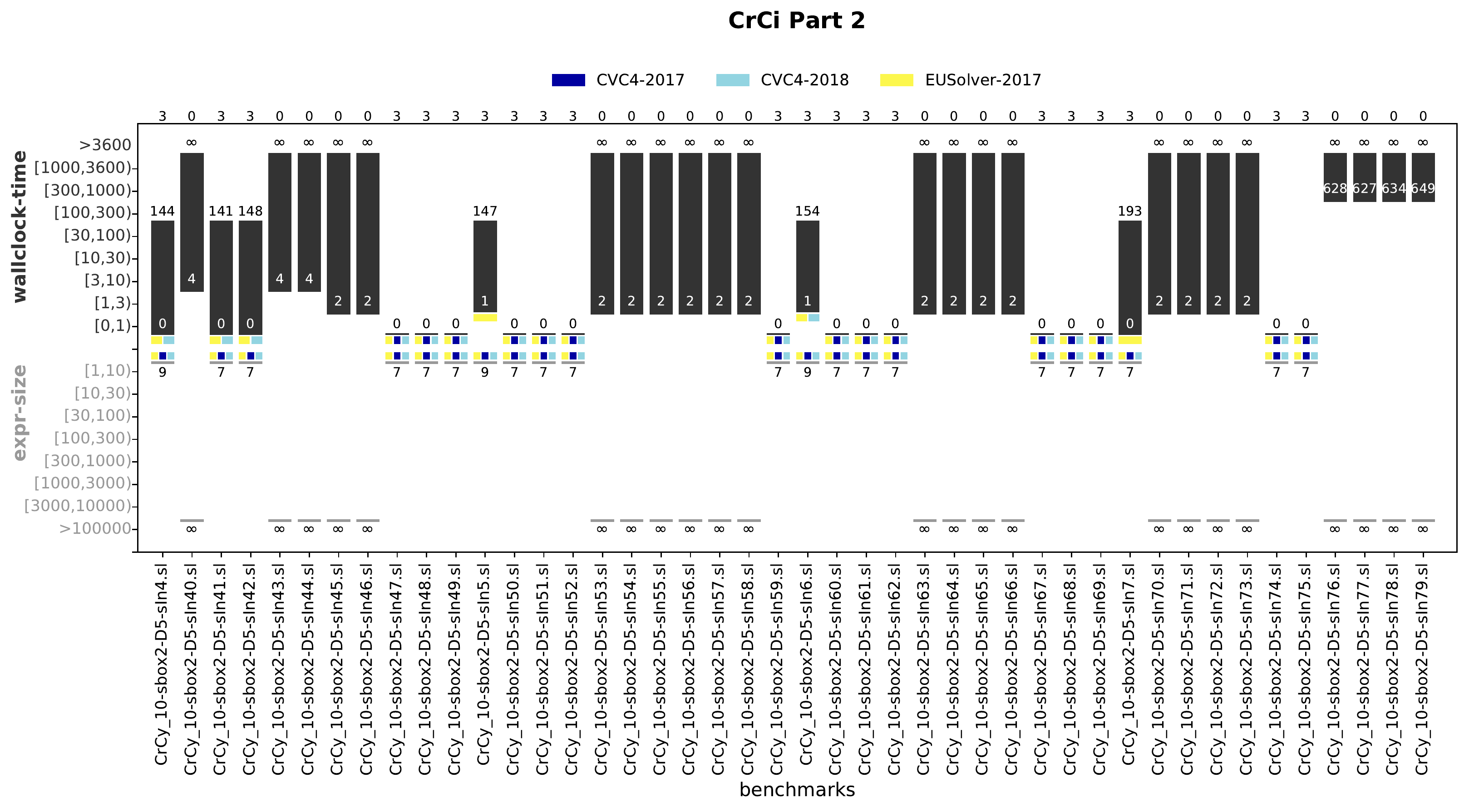}
			\end{tabular}
	}}
	\caption{Evaluation of crypto circuits category of the General track (Parts 1 \& 2).}
	\label{fig:crci-1}
\end{figure*}

\begin{figure*}
	\noindent\makebox[\textwidth]{
		\scalebox{0.625}{
			\begin{tabular}{c}
				\includegraphics[width=10in]{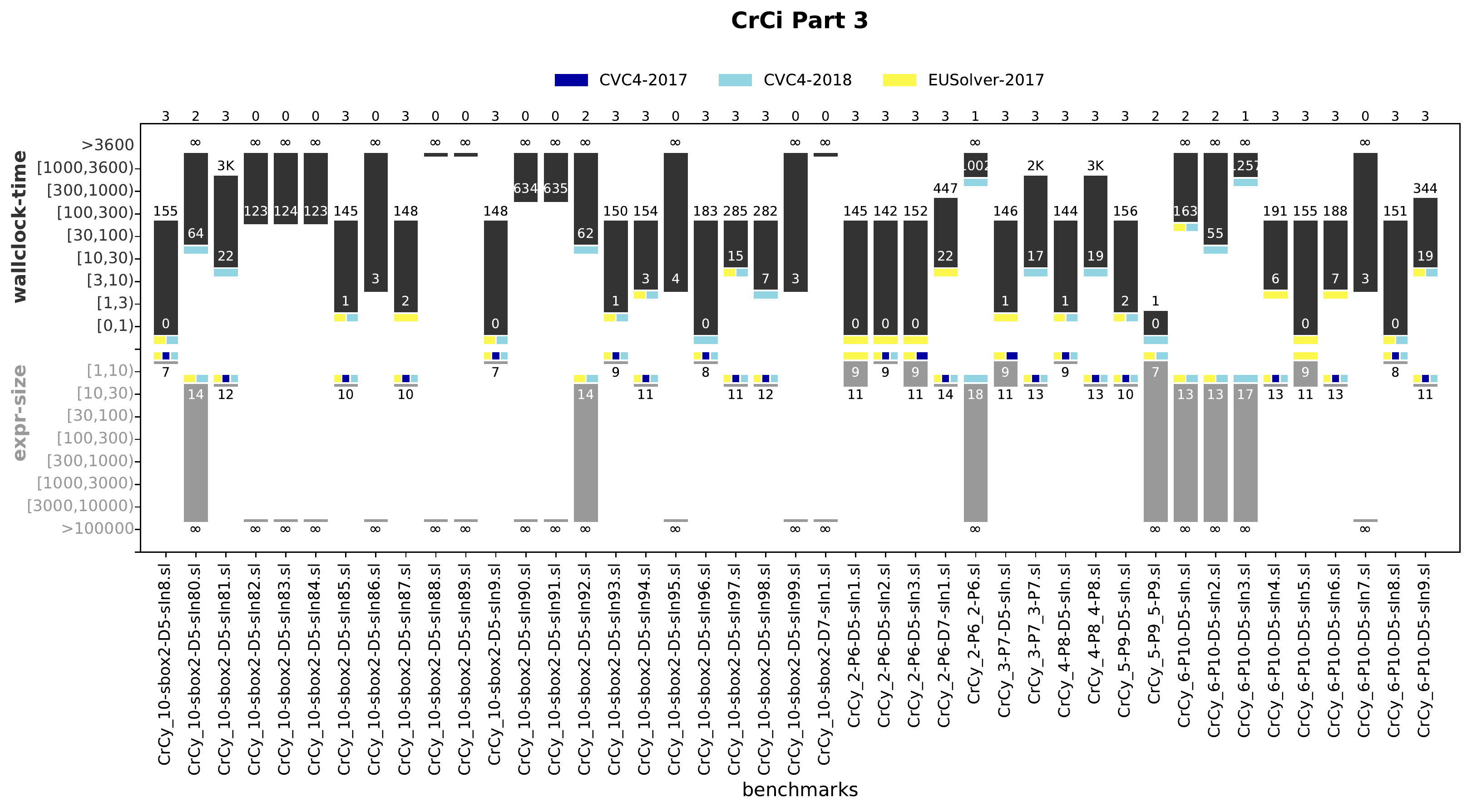} \\[3cm]
				\includegraphics[width=10in]{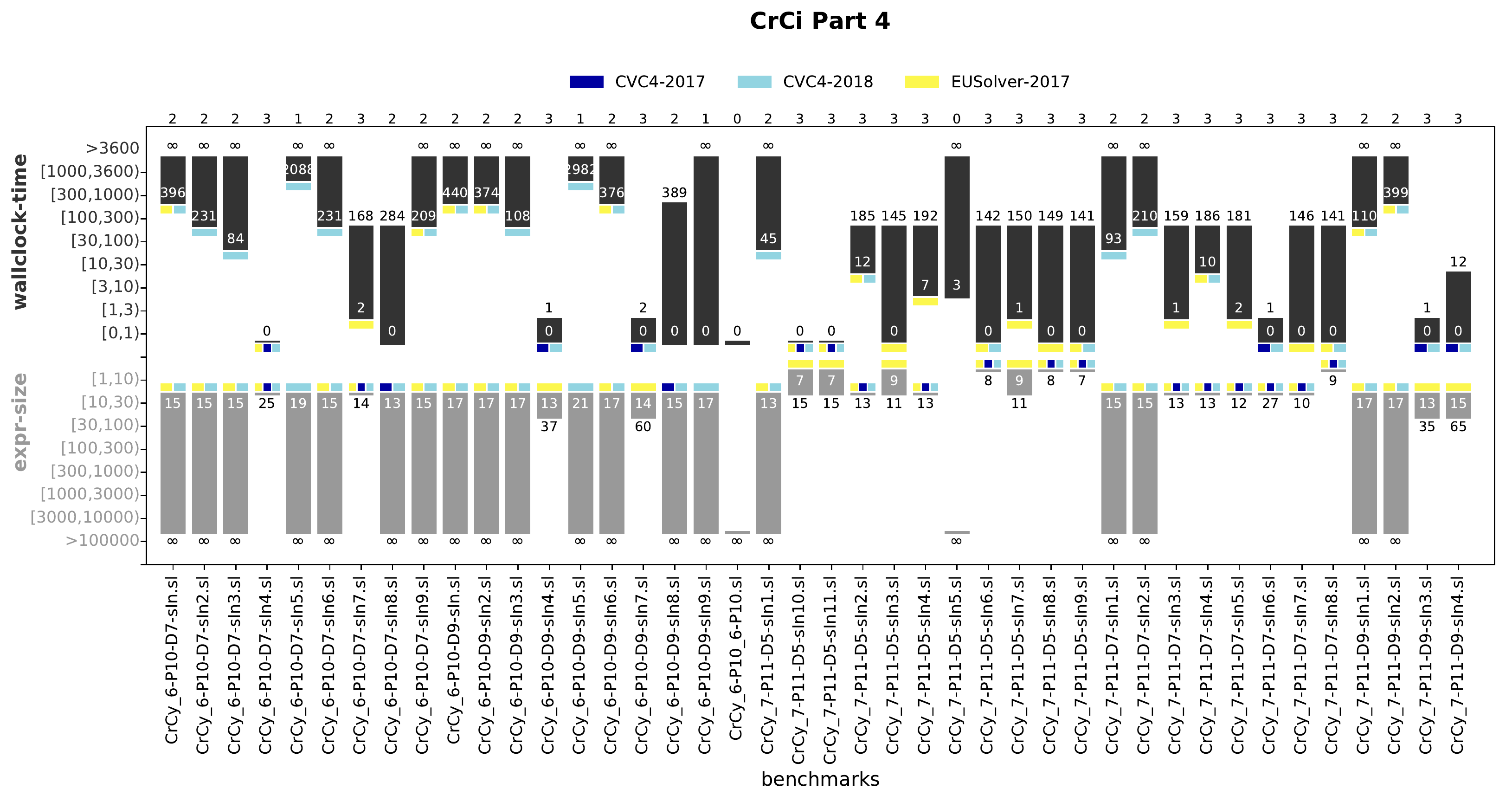}
			\end{tabular}
	}}
	\caption{Evaluation of crypto circuits category of the General track (Parts 3 \& 4).}
	\label{fig:crci-2}
\end{figure*}
	
\begin{figure*}
	\noindent\makebox[\textwidth]{
		\scalebox{0.625}{
			\begin{tabular}{c}
				\includegraphics[width=10in]{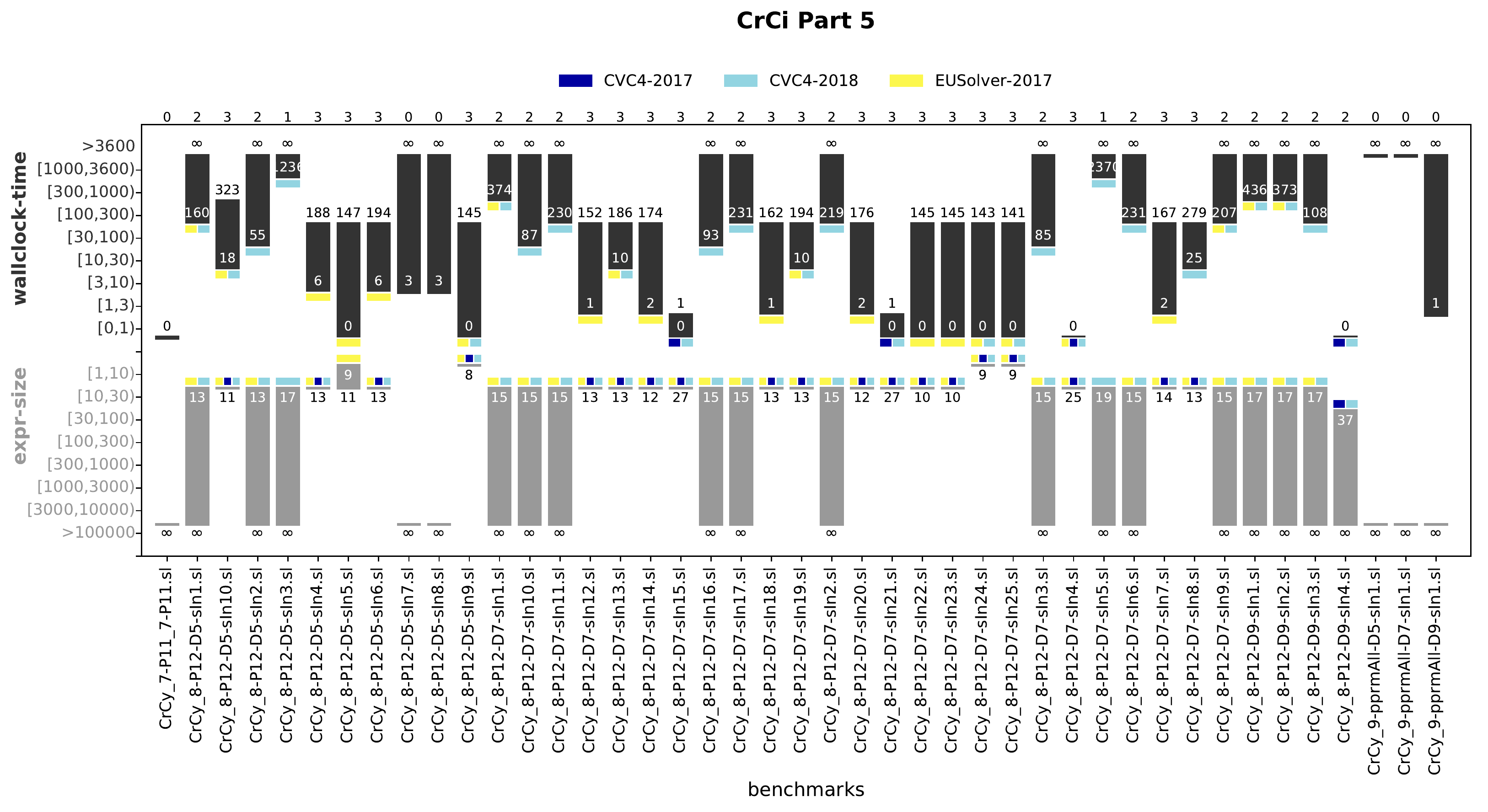}
			\end{tabular}
	}}
	\caption{Evaluation of crypto circuits category of the General track (Part 5).}
	\label{fig:crci-5}
\end{figure*}

\begin{figure*}
	\noindent\makebox[\textwidth]{
		\scalebox{0.6}{
			\begin{tabular}{c}
				\includegraphics[width=10in]{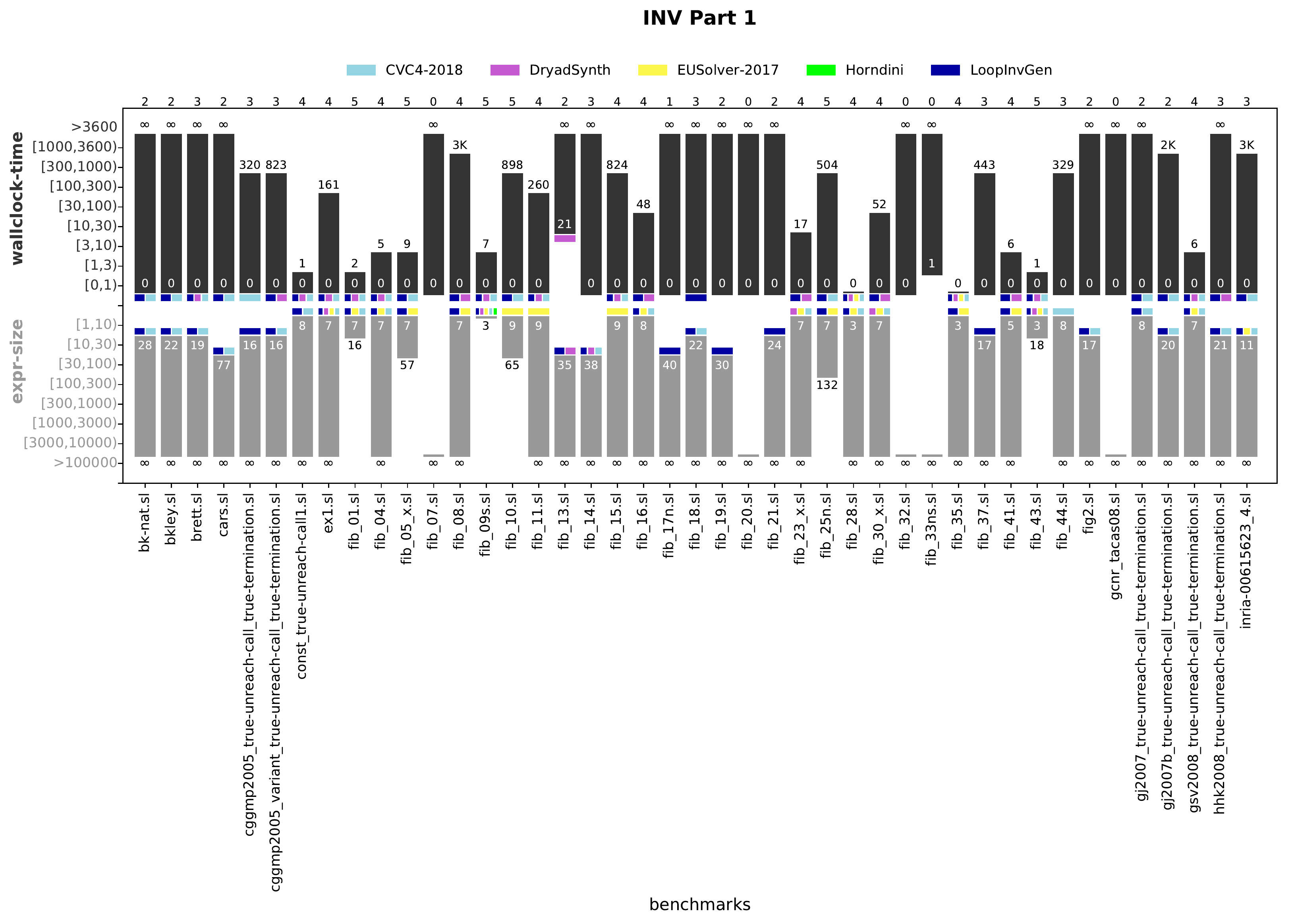}
			\end{tabular}
		}}
	\caption{Evaluation of Invariant track benchmarks (Part 1).}
	\label{fig:inv-results-1}
\end{figure*}
	
\begin{figure*}
	\noindent\makebox[\textwidth]{
		\scalebox{0.6}{
			\begin{tabular}{c}
				\includegraphics[width=10in]{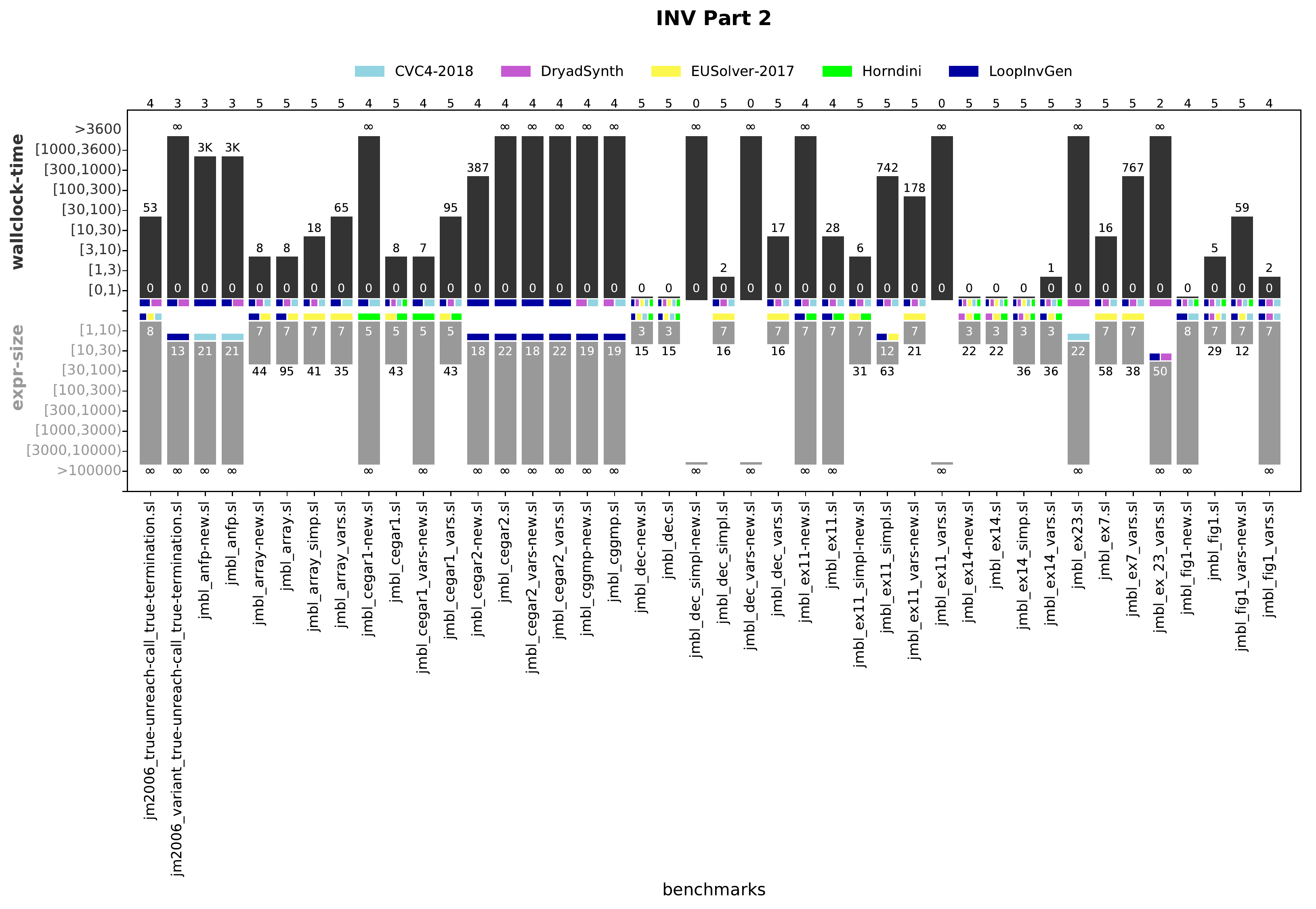} \\[5mm]
				\includegraphics[width=10in]{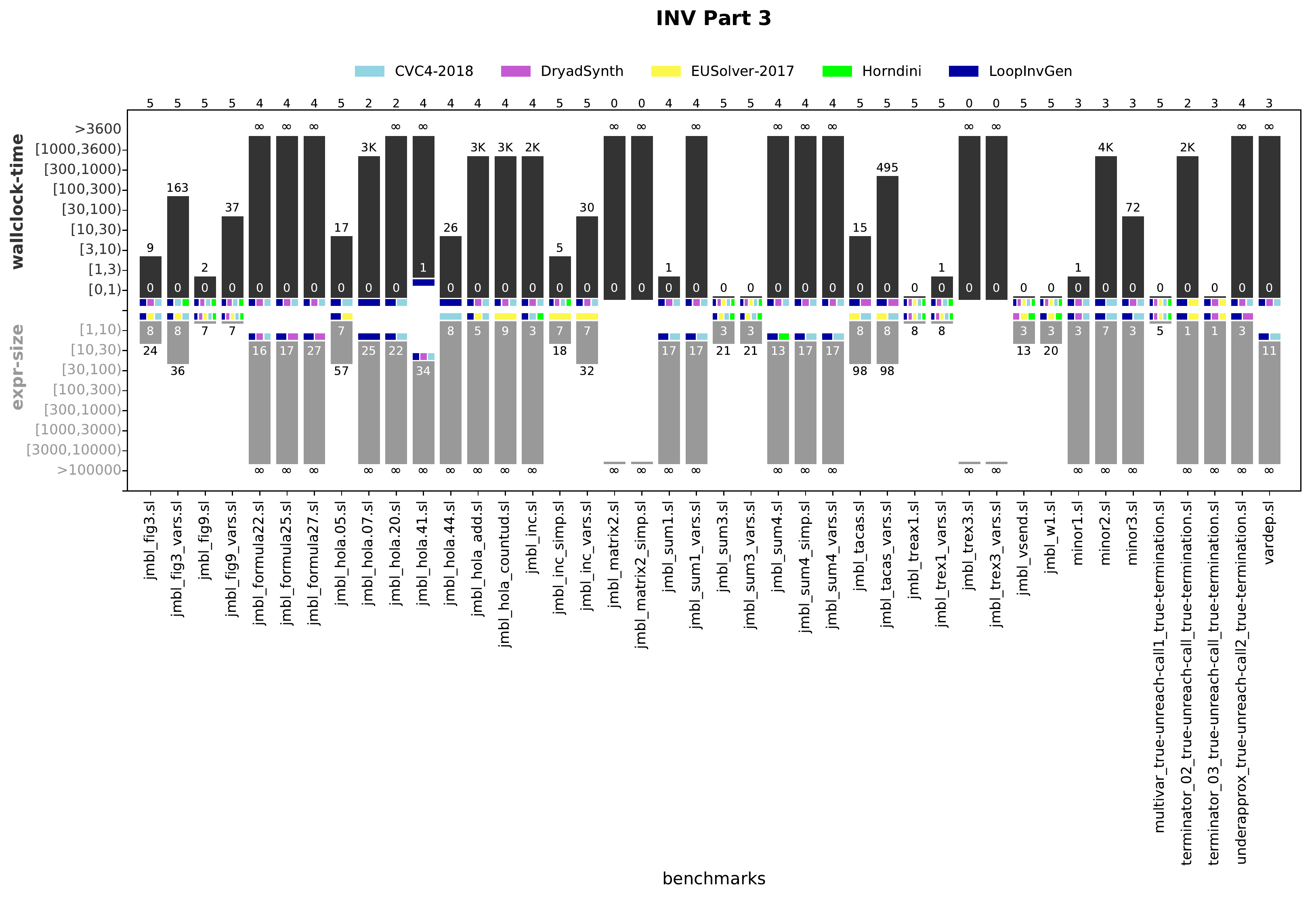}
			\end{tabular}
	}}
	\caption{Evaluation of Invariant track benchmarks (Parts 2 \& 3).}
	\label{fig:inv-results-2}
\end{figure*}

\begin{figure*}
	\noindent\makebox[\textwidth]{
		\scalebox{0.6}{
			\begin{tabular}{c}
				\includegraphics[width=10in]{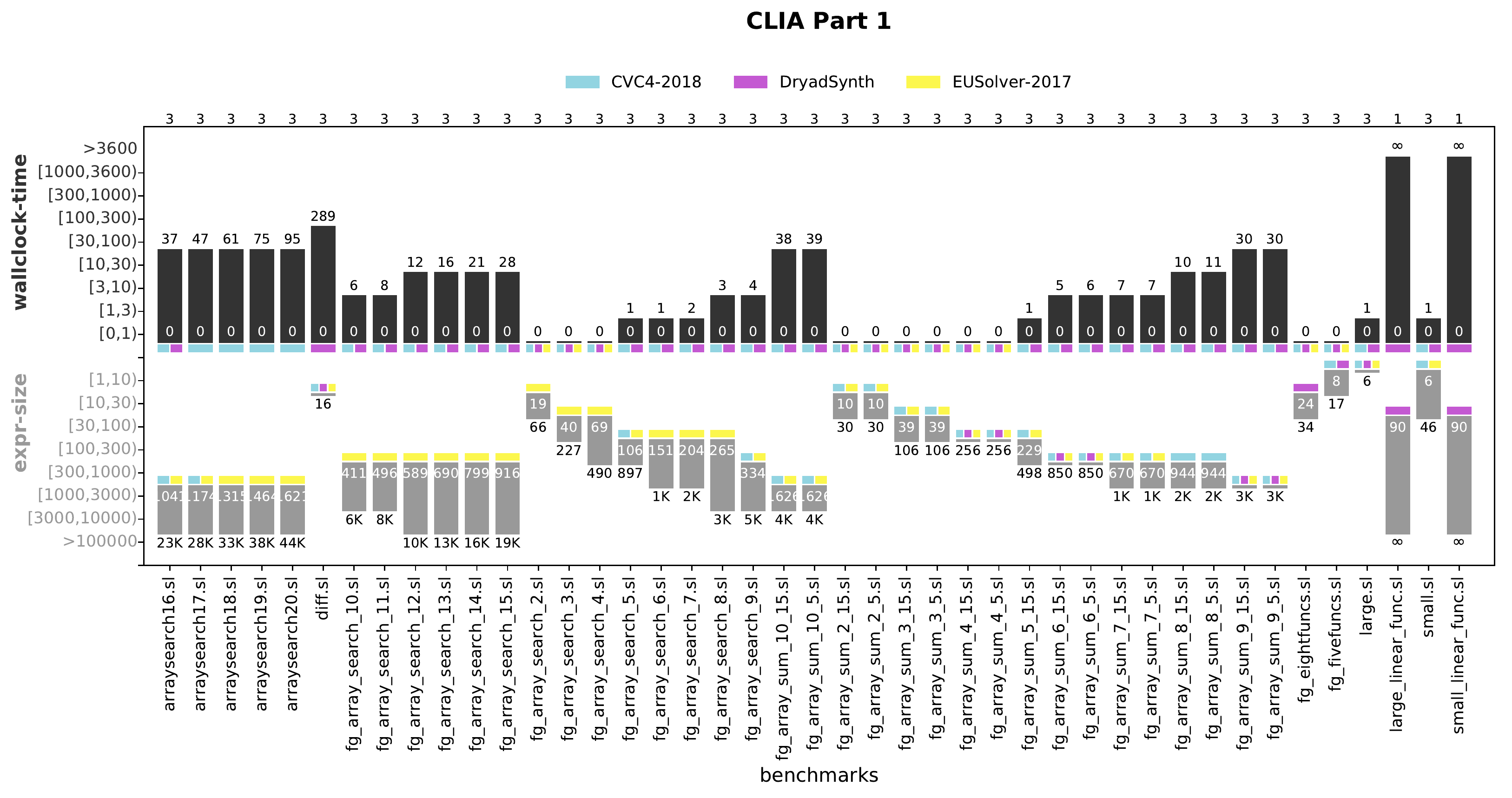} \\[3cm]
				\includegraphics[width=10in]{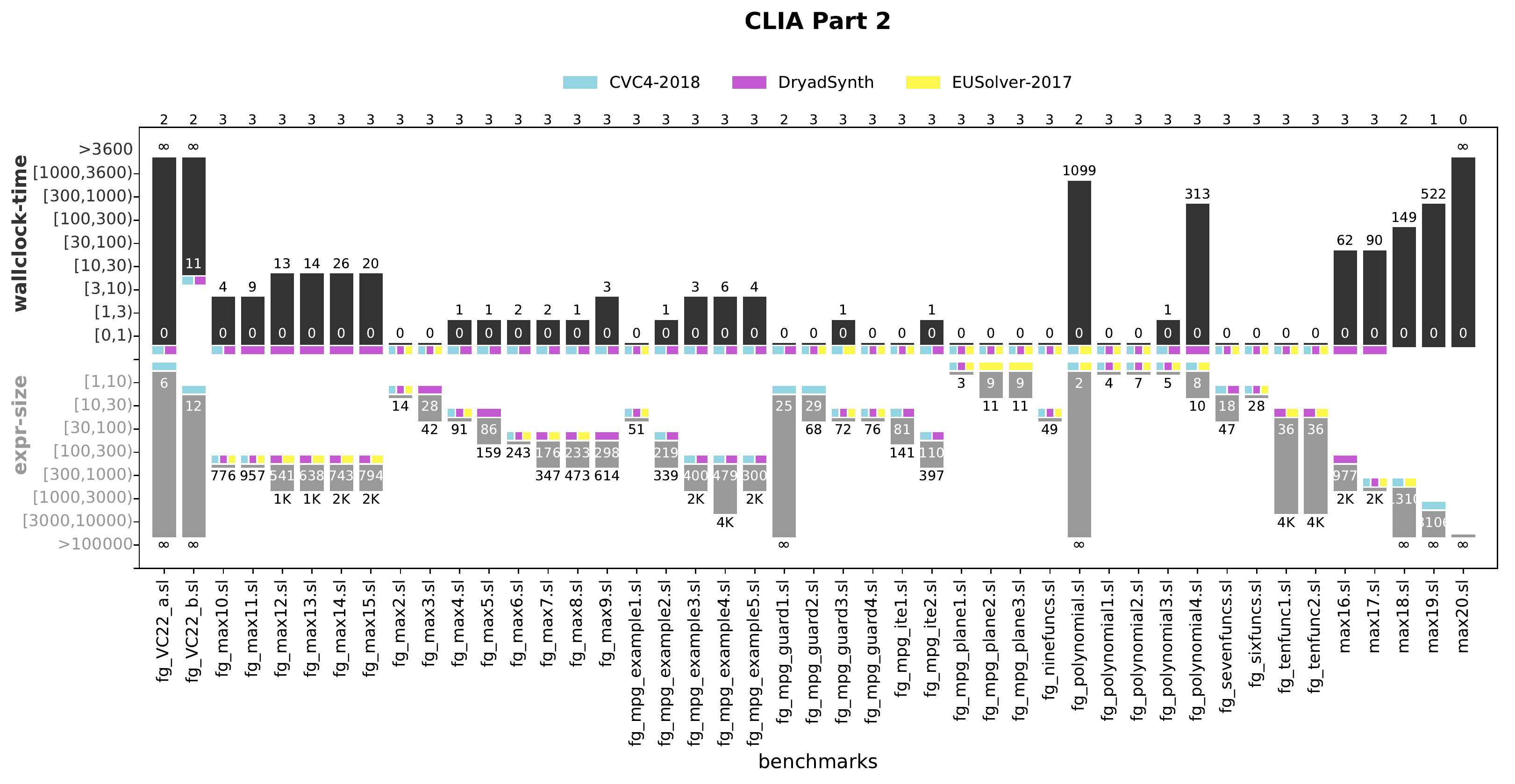} 
			\end{tabular}
	}}
	\caption{Evaluation of CLIA track benchmarks.}
	\label{fig:clia-results}
\end{figure*}

\begin{figure*}
	\noindent\makebox[\textwidth]{
		\scalebox{0.6}{
			\begin{tabular}{c}
				\includegraphics[width=10in]{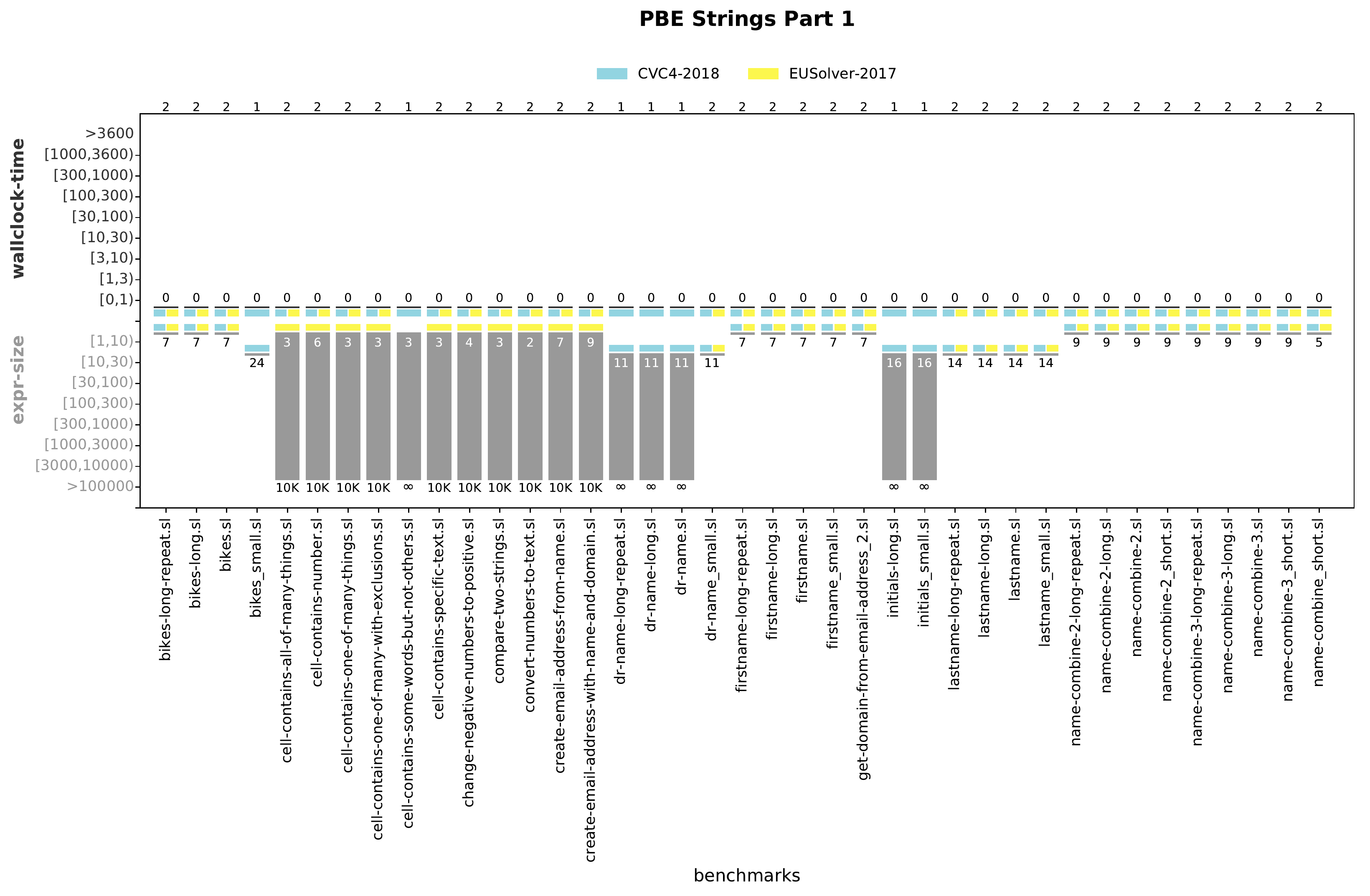} \\[3cm]
				\includegraphics[width=10in]{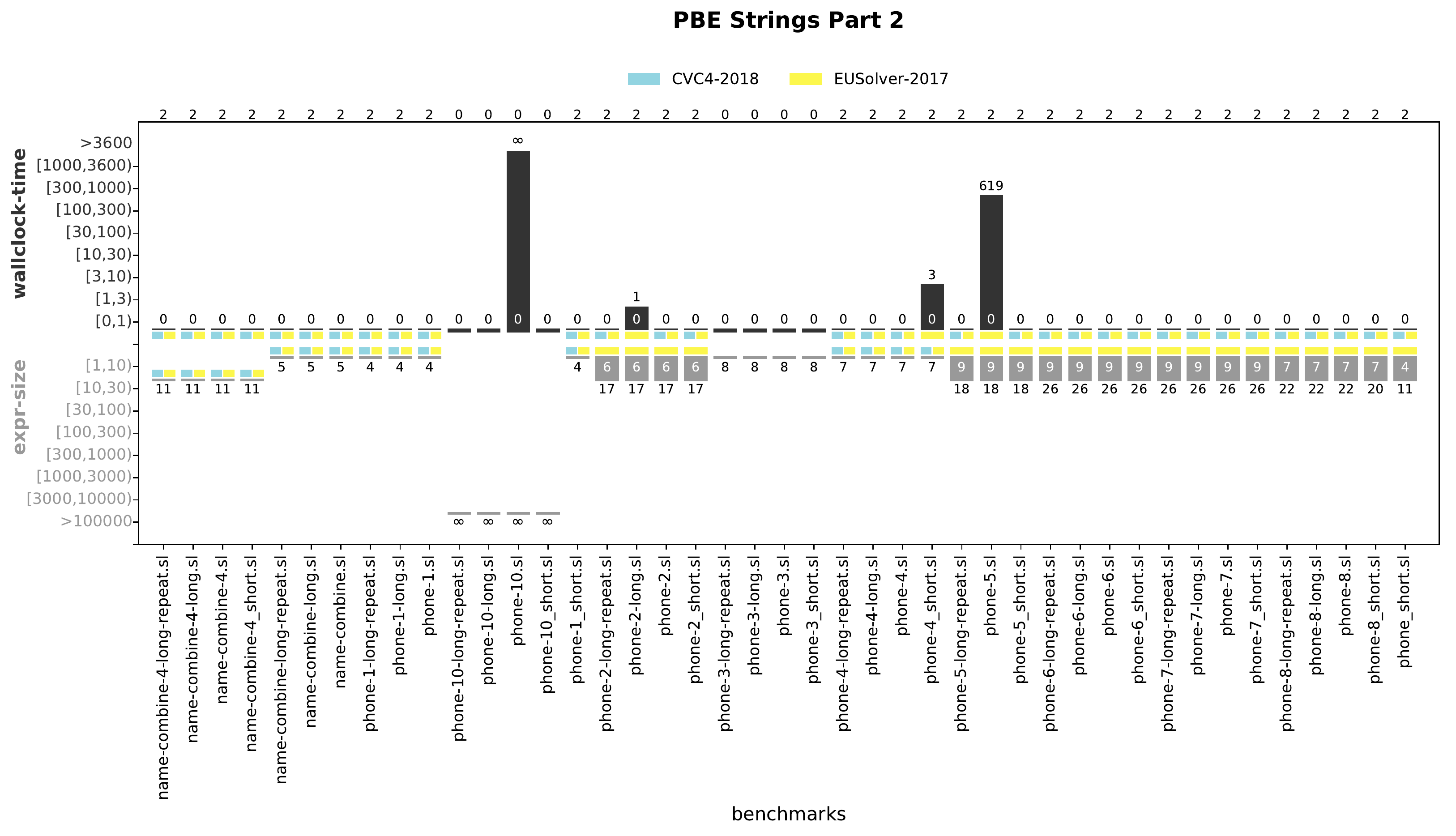} 
			\end{tabular}
		}}
	\caption{Evaluation of PBE Strings track benchmarks.}
	\label{fig:pbe-strings-results}
\end{figure*}

\section{Summary}
\label{sec:discussion}

This year's competition consisted of over $1600$ benchmarks, $107$ of which where contributed this year.
Five solvers competed this year, one of which was submitted by developers creating a tool for SyGuS-Comp for the first time.
All tools preformed remarkably well, on both existing and new benchmarks.
In particular, more than $74$\% of the current set of benchmarks from the general track are now solved.
However, there are several classes of problems that are still challenging for the current solvers,
especially the ``Instruction Selection'' and ``Multiple Functions'' categories;
and we hope the developers would continue to improve their SyGuS techniques and advance the state of the art.

\bibliographystyle{eptcs}
\bibliography{paper}

\end{document}